\documentclass[10pt]{iopart}
\usepackage{iopams,mathdots}
\newfont{\bfsf}{cmssbx10}
\newcommand\bfsfL{\mbox{\bfsf\symbol{'114}}}
\newcommand\bfsfM{\mbox{\bfsf\symbol{'115}}}

\newcommand\bfsfP{\mbox{\bfsf\symbol{'120}}}
\newcommand\bfsfU{\mbox{\bfsf\symbol{'125}}}
\newcommand\bfsfX{\mbox{\bfsf\symbol{'130}}}
\newcommand\bfsfeta{\mbox{\boldmath $\eta$}}
\newcommand\bfsfeps{\mbox{\boldmath $\epsilon$}}
\newcommand\bfsftau{\mbox{\boldmath $\tau$}}

\newcommand{\calL}{\mathcal{L}}

\newcommand{\triplen}{\mathinner{\mkern1mu\raise4.0pt%
\vbox{\kern4.0pt\hbox{.}}\mkern-5mu
\raise2.0pt\hbox{.}\mkern-5mu
\hbox{.}\mkern1mu}}
\begin{document}
\jl{1}

\title[Oscillating modes of a multicomponent ionic
mixture]{Damping and dispersion of oscillating modes of a
multicomponent ionic mixture in a magnetic field}
\author{G A Q Salvati and L G Suttorp}
\address{Instituut voor Theoretische Fysica, Universiteit van Amsterdam, 
Valckenierstraat 65, 1018 XE Amsterdam, The Netherlands}

\begin{abstract}
The collective-mode spectrum of a multicomponent magnetized ionic mixture
for small wave number $k$ is studied with the use of magnetohydrodynamics
and formal kinetic theory. Apart from the usual thermal and diffusive
modes, the spectrum contains a set of four oscillating modes. By evaluating
the $k^2$ contributions to the eigenfrequencies, the damping and the dispersion
of these oscillating modes are determined. The long-range nature of the
Coulomb interactions is shown to imply that Burnett terms with higher-order
gradients in the linear phenomenological laws have to be taken into account
in order to obtain a full description of all damping and dispersion
effects.
\end{abstract}


\vspace{5cm}


\vspace{5cm}

\maketitle

\section{Introduction}

The dynamical properties of classical Coulomb systems change considerably
when external magnetic fields are applied. In particular, the spectrum of
large-scale collective modes is strongly influenced by such fields. Whereas
the modes associated to the longitudinal and the transverse degrees of
freedom may be excited independently in an unmagnetized system, these
degrees of freedom get coupled when a magnetic field is switched on, at
least for modes propagating in an oblique direction with respect to the
magnetic field. As a consequence, both the well-known plasma oscillations
and the viscous modes get mixed when the isotropy of the system is lost by
applying a magnetic field. 

The influence of magnetic fields on the mode spectrum of Coulomb systems
has been studied in several model systems. For the one-component plasma
model (OCP), which consists of charged particles moving through an inert
neutralizing background of opposite charge, the changes in the mode
spectrum have been studied by means of magnetohydrodynamics \cite{LGSJSC85}
and in the framework of formal kinetic theory
\cite{LGSJSC85}--\cite{MCMTRKJRD87}. Since the dynamics of the magnetized
OCP model has an anomaly that is caused by a divergency in the heat
conductivity \cite{AJSLGS87,AJSLGS89}, a full treatment of the effects of a
magnetic field on the mode spectrum calls for a somewhat more general
model. The multicomponent ionic mixture (MIM) in a magnetic field is quite
useful in this respect \cite{AJSLGS90}.

In the magnetized MIM model several species of charged particles are
present. If the charge-mass ratios of the particles are not all equal, the
anomalous behaviour of the heat transport is no longer present, as has been
shown in Ref.\ \cite{AJSLGS90}. In that paper the modes of the magnetized MIM
have been studied by formal kinetic theory. More recently \cite{GAQSLGS01}, a
magnetohydrodynamical approach has been used to determine the influence
of the magnetic field on the modes. The results were found to corroborate
those found in \cite{AJSLGS90}. In the long-wavelength limit the spectrum of
collective modes consists of four oscillating modes with a non-vanishing
frequency, and of a set of coupled heat and diffusion modes with
frequencies tending to 0 for vanishing wave number. By including terms of
higher order in the wave number, the latter modes could be
discerned. The second-order terms in the mode frequencies describe
damping that is caused by dissipative transport phenomena. 

In Ref.\ \cite{AJSLGS90} and \cite{GAQSLGS01} the mode frequencies of the
oscillating modes have been obtained in the limit of vanishing wave number
only. Hence, dispersion effects for these modes could not be
studied. Furthermore, damping effects that are of second order in the
wave number were not treated either. In the present paper we wish to
fill this gap by evaluating the second-order contributions to the
oscillating-mode frequencies in a systematic way. 

As in our previous paper \cite{GAQSLGS01} we shall use both a
magnetohydrodynamical approach and formal kinetic theory. Interestingly
enough, it turns out that standard magnetohydrodynamics is not enough to
determine the frequencies of the oscillating modes in second order of the
wave number. As we will show, so-called Burnett terms will be needed to
obtain a consistent picture. For systems of neutral particles these Burnett
terms \cite{DB36}--\cite{JJB81}, which describe non-local effects in
transport phenomena, are of rather limited importance. They play a role
only when one is interested in contributions to mode frequencies that are
of higher than second order in the wave number. For charged-particle
systems, however, the long-range nature of the Coulomb interaction leads to
a reshuffling of terms in the magnetohydrodynamical equations, which
enhances the relevance of the Burnett contributions.

The paper is organized as follows. In Section \ref{sec2} the conservation
laws and the balance equations of linearized magnetohydrodynamics for the
magnetized MIM model are formulated.  The linear phenomenological laws
relating thermodynamic forces and flows are presented. In particular, the
Burnett contributions will be written down explicitly.  The
Fourier-transformed versions of these magnetohydrodynamical equations are
given in Section \ref{sec3}. The next two sections contain the analysis
that leads to our main result, namely the eigenfrequencies of the
oscillating modes up to second order in the wavelength. In Sections
\ref{sec6} and \ref{sec7} a kinetic approach is used to rederive these
frequencies and to obtain the kinetic equivalents of the transport
coefficients, both in leading order and in the higher-order Burnett
regime. A few auxiliary thermodynamical relations and the basic formalism
of the kinetic eigenvalue problem are presented in two appendices.

\section{Conservation laws, balance equations and phenomenological relations}
\setcounter{equation}{0}  
\label{sec2}
An ionic mixture consists of several species of charged particles (with
charges of the same sign), which move in an inert uniform neutralizing
background. If a uniform external magnetostatic field is present as well,
the motion of each of the particles is governed by the electrostatic forces
from the other particles and from the background, and by the Lorentz force
due to the external field. We assume that the motion of the particles is so
slow that the magnetic fields generated by them can be disregarded.

In thermal equilibrium the particle density $n_\sigma$ of species $\sigma$
(with $\sigma=1,\ldots,s$) is uniform and time-independent. The total
charge density $q_v=\sum_\sigma e_\sigma n_\sigma$ (with $e_\sigma$ the
charge of a particle of species $\sigma$) is matched by the background
charge density $-q_v$. The total mass density $m_v$ follows as $\sum_\sigma
m_\sigma n_\sigma$, with $m_\sigma$ the mass of a particle of species
$\sigma$.

If equilibrium is slightly perturbed, the particle densities acquire
deviations $\delta n_\sigma(\bi{r},t)$ that in general are non-uniform and
time-dependent. Moreover, each species in the perturbed system will get a
non-uniform and time-dependent velocity $\bi{v}_{\sigma}(\bi{r},t)$. In the
following we assume that the perturbations vary slowly in space and
time. The characteristic length $l$ of the spatial variations should be
large compared to the average interparticle distance $n_{\sigma}^{-1/3}$,
for all $\sigma$. The characteristic time for temporal variations should be
large compared to $l/c$, with $c$ the velocity of light.

The continuity equation for species $\sigma$ reads in first order of the
perturbations \cite{GAQSLGS01}:
\begin{equation}
\label{2.1}
\frac{\partial}{\partial t}\, \delta n_{\sigma}(\bi{r},t) = 
-n_{\sigma}\nabla\cdot \bi{v}(\bi{r},t)
-\frac{1}{m_{\sigma}}\nabla\cdot\bi{J}_{\sigma}(\bi{r},t)
\end{equation}
Here $\bi{v}(\bi{r},t)=\sum_{\sigma} m_{\sigma}\, n_{\sigma}\,
\bi{v}_{\sigma}(\bi{r},t)/m_v$ is the barycentric velocity and
$\bi{J}_{\sigma}=m_{\sigma}\, n_{\sigma}
[\bi{v}_{\sigma}(\bi{r},t)-\bi{v}(\bi{r},t)]$ the diffusion flow of species
$\sigma$ with respect to the barycentric motion. These flows satisfy the
identity $\sum_{\sigma} \bi{J}_{\sigma}(\bi{r},t)=0$. Upon multiplying by
$e_\sigma$ and summing over $\sigma$ we get the conservation law for the
total charge:
\begin{equation} 
\label{2.2}   
\frac{\partial}{\partial t}\, \delta q_v(\bi{r},t) = -q_v\nabla\cdot 
\bi{v}(\bi{r},t)-\sum_{\sigma(\neq 1)}\; 
\left(\frac{e_{\sigma}}{m_{\sigma}}-\frac{e_1}{m_1}\right)
\, \nabla\cdot\bi{J}_{\sigma}(\bi{r},t)
\end{equation}    
As an independent set we will choose (\ref{2.2}) in combination with
(\ref{2.1}) for $\sigma\neq 1$.

The time dependence of the barycentric motion is given by the equation of
motion, which up to first order in the perturbations reads \cite{GAQSLGS01}:
\begin{eqnarray}
\label{2.3}
\fl  m_v\frac{\partial}{\partial t}\, \bi{v}(\bi{r},t) = 
-\nabla\delta p(\bi{r},t)-\nabla\cdot\delta\bfsfP(\bi{r},t)
+q_v\left[\bi{E}(\bi{r},t)
+c^{-1}\bi{v}(\bi{r},t)\wedge\bi{B}\right]\nonumber\\
+c^{-1} \sum_{\sigma(\neq 1)}\left(\frac{e_{\sigma}}{m_{\sigma}}-
\frac{e_1}{m_1}\right)\bi{J}_{\sigma}(\bi{r},t)\wedge\bi{B}
\end{eqnarray}
Here $\delta p$ is the perturbation of the (local) equilibrium pressure,
while $\delta\bfsfP$ is the viscous pressure tensor. The electric field is
determined by $\delta q_v$ through Gauss's law
$\nabla\cdot\bi{E}(\bi{r},t)=\delta q_v(\bi{r},t)$. Furthermore, $\bi{B}$
is the uniform external magnetostatic field.

The perturbation $\delta\varepsilon$ of the internal energy 
density satisfies a balance equation, which gets the form of a conservation
law when only terms up to first order in the perturbations are taken into account:
\begin{equation}
\label{2.4}
\frac{\partial}{\partial t}\, \delta\varepsilon(\bi{r},t) =
-h_v\nabla\cdot\bi{v}(\bi{r},t)
-\nabla\cdot\bi{J}_{\varepsilon}(\bi{r},t)
\end{equation}
Here $h_v=\varepsilon+p$ is the enthalpy per volume, with $\varepsilon$ the
(uniform) equilibrium internal energy density and $p$ the equilibrium
pressure. Furthermore $\bi{J}_{\varepsilon}$ is the heat flow.

The conservation laws and the equation of motion contain the heat flow
$\bi{J}_{\varepsilon}$, the diffusion flows $\bi{J}_{\sigma}$ and the
viscous pressure tensor $\delta\bfsfP$. These quantities can be expressed
in terms of thermodynamic forces by means of phenomenological relations
\cite{dGM62}. The forces that drive the heat flow and the diffusion flows
are \cite{GAQSLGS01}
\begin{eqnarray}
\label{2.5}
\bi{X}_{\varepsilon}=\nabla\left(\frac{1}{T}\right) \\
\label{2.6}
\bi{X}_{\sigma}
=-\frac{1}{T}\left[\nabla\left(\frac{\mu_{\sigma}}{m_{\sigma}}-
\frac{\mu_1}{m_1}\right)\right]_T 
+\frac{1}{T}\left(\frac{e_{\sigma}}{m_{\sigma}}-\frac{e_1}{m_1}\right)
(\bi{E}+\frac{1}{c}\bi{v}\wedge\bi{B})
\end{eqnarray}
The subscript ${}_T$ indicates that the gradients of the chemical
potentials $\mu_\sigma$ have to be taken at constant temperature $T$.

In the standard formalism of non-equilibrium thermodynamics the diffusion
flows are written as linear combinations of $\bi{X}_{\varepsilon}$ and
$\bi{X}_{\sigma}$:
\begin{equation}
\label{2.7}
\bi{J}_{\sigma}=
\bfsfL_{\sigma\varepsilon}\cdot\bi{X}_{\varepsilon} 
+\sum_{\sigma'(\neq 1)} \bfsfL_{\sigma\sigma'}\cdot\bi{X}_{\sigma'}
\end{equation}
The phenomenological coefficients $\bfsfL_{\sigma\varepsilon}$ and
$\bfsfL_{\sigma\sigma'}$ are anisotropic tensors of second rank. 

The modified heat flow $\bi{J}'_{\varepsilon}=\bi{J}_{\varepsilon}-
\sum_{\sigma}(h_{\sigma}/m_{\sigma})\bi{J}_{\sigma}$, with $h_{\sigma}$ the
partial specific enthalpy per particle, is a linear combination of the
thermodynamic forces $\bi{X}_{\varepsilon}$ and $\bi{X}_{\sigma}$:
\begin{equation}
\label{2.8}
\bi{J}'_{\varepsilon}=
\bfsfL_{\varepsilon\varepsilon}\cdot \bi{X}_{\varepsilon}
+\sum_{\sigma(\neq 1)} \bfsfL_{\varepsilon\sigma}\cdot\bi{X}_{\sigma}
\end{equation}
As a consequence, the heat flow $\bi{J}_{\varepsilon}$
itself gets the form \cite{GAQSLGS01}:
\begin{equation}
\label{2.9}
\bi{J}_{\varepsilon}=
\bar{\bfsfL}_{\varepsilon\varepsilon}\cdot \bi{X}_{\varepsilon}+
\sum_{\sigma(\neq 1)} \bar{\bfsfL}_{\varepsilon\sigma}\cdot\bi{X}_{\sigma}
\end{equation}
with the tensorial coefficients
\begin{eqnarray}
\label{2.10}
\bar{\bfsfL}_{\varepsilon\varepsilon}=\bfsfL_{\varepsilon\varepsilon}+
\sum_{\sigma'(\neq 1)}\left(\frac{h_{\sigma'}}{m_{\sigma'}}-
\frac{h_1}{m_1}\right)\bfsfL_{\sigma'\varepsilon}\\
\label{2.11}
\bar{\bfsfL}_{\varepsilon\sigma}=\bfsfL_{\varepsilon\sigma}+
\sum_{\sigma'(\neq 1)}\left(\frac{h_{\sigma'}}{m_{\sigma'}}-
\frac{h_1}{m_1}\right)\bfsfL_{\sigma'\sigma}
\end{eqnarray}
The tensors $\bfsfL$ satisfy the Onsager relations:
\begin{equation}
\label{2.12}
\fl L^{ij}_{\varepsilon\varepsilon}(\bi{B})
=L^{ji}_{\varepsilon\varepsilon}(-\bi{B}) \quad , \quad 
L^{ij}_{\varepsilon\sigma}(\bi{B})
=L^{ji}_{\sigma\varepsilon}(-\bi{B}) \quad , \quad
L^{ij}_{\sigma\sigma'}(\bi{B})=L^{ji}_{\sigma'\sigma}(-\bi{B}) 
\end{equation}
Similar relations hold for $\bar{\bfsfL}$.

Finally, in standard non-equilibrium thermodynamics the viscous pressure
tensor is proportional to the thermodynamic force $\bfsfX_v=-\nabla\bi{v}$:
\begin{equation}
\label{2.13}
\delta\bfsfP= \bfsfeta : \bfsfX_v
\end{equation}
The symbol $:$ denotes a double contraction, so that the $ij$-component of
the right-hand side is $\eta^{ijmn} X^{mn}_v$.  The components of the
fourth-rank anisotropic viscosity tensor $\bfsfeta$ satisfy the Onsager
relations:
\begin{equation}
\label{2.14}
\eta^{ijmn}(\bi{B})=\eta^{mnij}(-\bi{B})
\end{equation}
Furthermore, $\eta^{ijmn}$ is symmetric in $ij$ and in $mn$. 

For neutral-particle systems the above phenomenological relations are
sufficiently general to determine the damping and dispersion of the
collective modes up to second order in the wave number. However, for a
system of charged particles this need not be true any longer, at least when
various species of particles are present. Whereas the conventional
phenomenological relations are still adequate to determine the modes of the
OCP model \cite{LGSJSC85}--\cite{MCMTRKJRD87}, the standard formalism has
to be generalized for the case of the MIM model. For that system the
damping and dispersion of the diffusion and heat modes up to second order
in the wave number are correctly found by inserting the above
phenomenological relation in the balance equations, as shown in
\cite{GAQSLGS01}. In contrast, one has to include Burnett terms with
higher-order gradient operators in the phenomenological relations in order
to determine the damping and dispersion of the oscillating modes up to
second order in the wave number. 

The relevance of higher-order Burnett terms becomes obvious upon inspecting
the charge conservation law. When (\ref{2.7}) with (\ref{2.6}) is
substituted in the last term of (\ref{2.2}), the divergence of the
diffusion flow $\bi{J}_{\sigma}$ leads to a spatial derivative of the
electric field. The long-range nature of the Coulomb field implies that
such a derivative does not necessarily yield a contribution of first order
in the wavenumber in Fourier language. In fact, from Gauss's law it follows
that the spatial derivative of the electric field is proportional to
$\delta q_v$ itself, so that it is of the same order in the wave number as
the left-hand side of (\ref{2.2}). Hence, if one wishes to determine all
contributions to the oscillating-mode frequencies up to second order in the
wavenumber, one has to go further than the standard linear law
(\ref{2.7}). A similar effect is seen in the equation of motion
(\ref{2.3}), of which the last term is proportional to the diffusion
flows. According to (\ref{2.6}) and (\ref{2.7}) the latter contain
electrodynamic terms without gradient operators, so that in Fourier
language terms appear that are of less than first order in the wavenumber.

From both these arguments it becomes clear that in order to determine the
damping and dispersion of the collective modes up to second order in the
wave number one is forced to generalize the phenomenological expression for
the diffusion flow by including Burnett terms of higher order in the
gradients. By invoking parity invariance one finds the following general
form of the diffusion flow up to second order in the gradients:
\begin{equation}
\label{2.15}
\bi{J}_{\sigma}=
\bfsfL_{\sigma\varepsilon}\cdot\bi{X}_{\varepsilon} 
+\sum_{\sigma'(\neq 1)} \bfsfL_{\sigma\sigma'}\cdot\bi{X}_{\sigma'}
+\sum_{\sigma'(\neq 1)} \bfsfL^{\rm (c)}_{\sigma\sigma'}\;\triplen\;
\bfsfX^{\rm (c)}_{\sigma'}
+\bfsfL^{\rm (c)}_{\sigma v}\;\triplen\;\bfsfX^{\rm (c)}_{v}
\end{equation}
with the generalized thermodynamic forces:
\begin{eqnarray}
\label{2.16}
\bfsfX^{\rm (c)}_{\sigma}&=&
\frac{1}{T}\left(\frac{e_{\sigma}}{m_{\sigma}}-\frac{e_1}{m_1}\right)
\nabla\nabla(\bi{E}+\frac{1}{c}\bi{v}\wedge\bi{B}) \\
\label{2.17}
\bfsfX^{\rm (c)}_{v}&=&-\frac{1}{T}\nabla\nabla\bi{v}
\end{eqnarray}
Both $\bfsfL^{\rm (c)}_{\sigma\sigma'}$ and $\bfsfL^{\rm (c)}_{\sigma v}$
are fourth-rank tensors, with the following symmetry properties: $L^{{\rm
(c)}ijmn}_{\sigma\sigma'}$ is symmetric in $jm$ , while $L^{{\rm
(c)}ijmn}_{\sigma v}$ is symmetric in its last three indices
$jmn$. Furthermore, $\bfsfL^{\rm (c)}_{\sigma\sigma'}$ has the Onsager
symmetry:
\begin{equation}
\label{2.18}
 L^{{\rm (c)}ijmn}_{\sigma\sigma'}(\bi{B})= L^{{\rm (c)}njmi}_{\sigma'\sigma}(-\bi{B})
\end{equation}
The symbol $\triplen$ in (\ref{2.15}) indicates a triple contraction; the
$i$-component of the last term, for instance, is $L^{{\rm (c)}ijmn}_{\sigma
v}X^{{\rm (c)}jmn}_v$.

According to (\ref{2.15}) with (\ref{2.17}), the diffusion flow depends on
the gradient of the thermodynamic force $\bfsfX_v$. Likewise, the viscous
pressure gets coupled to the gradient of the force $\bi{X}_{\sigma}$ when
higher-order terms in the gradients are included:
\begin{equation}
\label{2.19}
\delta\bfsfP= \bfsfeta : \bfsfX_v + \sum_{\sigma(\neq 1)} \bfsfL^{\rm (c)}_{v\sigma}
: \bfsfX^{\rm (c')}_{\sigma}
\end{equation}
with the generalized thermodynamic force
\begin{equation}
\label{2.20}
\bfsfX^{\rm (c')}_{\sigma}=
\frac{1}{T}\left(\frac{e_{\sigma}}{m_{\sigma}}-\frac{e_1}{m_1}\right)\nabla
(\bi{E}+\frac{1}{c}\bi{v}\wedge\bi{B})
\end{equation}
and a fourth-tank tensorial coefficient $\bfsfL^{\rm (c)}_{v\sigma}$. It is
related to $\bfsfL^{\rm (c)}_{\sigma v}$ by the Onsager relation 
\begin{equation}
\label{2.21}
 L^{{\rm (c)}ijmn}_{v\sigma}(\bi{B})= L^{{\rm (c)}njmi}_{\sigma v}(-\bi{B})
\end{equation}
which is similar to (\ref{2.18}). As a consequence, 
$ L^{{\rm (c)}ijmn}_{v\sigma}$ is symmetric in $ijm$.

The generalized phenomenological relations (\ref{2.15}) and (\ref{2.19})
contain terms with higher-order gradients of the charge density $\delta
q_v$ and the barycentric velocity $\bi{v}$. These terms represent non-local
effects in the phenomenological relations. Clearly, the diffusion flow and
the viscous pressure at a certain position are partly determined by the
charge density and the velocity in the neighbourhood of that position. The
relevance of such non-local effects need not surprise for an ionic mixture
in which long-range interparticle interactions are present.

\section{Fourier transforms of the balance equations}\label{sec3}
\setcounter{equation}{0}

To determine the properties of the collective modes we need the Fourier
transforms of the balance equations. The Fourier transform of the particle 
conservation law (\ref{2.1}) follows upon substitution of
(\ref{2.15}), with (\ref{2.5}), (\ref{2.6}), (\ref{2.16}) and
(\ref{2.17}). Up to second order in the wave number we get
\begin{eqnarray}
\label{3.1}
\fl \frac{\partial}{\partial t} \, \delta n_{\sigma}=
-\rmi\,k\,\omega_{p}\,\frac{n_{\sigma}}{q_v}\,\sqrt{m_v}\,\hat\bi{k}\cdot\bi{v}\,
-k\,\frac{1}{m_{\sigma}\,T}\,
\hat\bi{k}\cdot\bfsfL_{\sigma}\cdot\left(
\hat\bi{k}\,\frac{\delta q_v}{k} 
+\rmi\,\gamma \,\sqrt{m_v} \,\bi{v}\wedge\hat\bi{B}\right)\nonumber\\
+k^2 \,\frac{1}{m_{\sigma}}\,
\hat\bi{k}\cdot\bfsfL_{\sigma\varepsilon}\cdot\hat\bi{k}\,
 \delta\left(\frac{1}{T}\right)
\nonumber\\
-k^2 \, \frac{1}{m_{\sigma}\,T}\sum_{\sigma'(\neq 1)} \, 
\hat\bi{k}\cdot\bfsfL_{\sigma\sigma'}\cdot\hat\bi{k}\,
\left(\frac{\delta\mu_{\sigma'}}{m_{\sigma'}}-
\frac{\delta\mu_1}{m_1}\right)_T 
\end{eqnarray}
Here $\hat\bi{k}=\bi{k}/k$ is a unit vector in the direction of the
wave vector. Furthermore, we introduced the following combination of
phenomenological coefficients:
\begin{equation}
\label{3.2}
\bfsfL_{\sigma}=\sum_{\sigma'(\neq 1)}
\left(\frac{e_{\sigma'}}{m_{\sigma'}}
-\frac{e_1}{m_1}\right)\, \bfsfL_{\sigma\sigma'}
\end{equation}
The ratio of the plasma frequency $\omega_p=q_v/\sqrt{m_v}$ and the
cyclotron frequency $\omega_c=q_vB/(m_vc)$ is denoted as
$\gamma=\omega_c/\omega_p$. As a consequence of the long range of the
Coulomb interactions, large-scale charge perturbations are suppressed
effectively. In fact, one may prove \cite{AJVWLGS87} that in thermal
equilibrium the Fourier-transformed fluctuations $\delta q_v$ are
proportional to $k$. For that reason, the `reduced' charge-density
perturbation $\delta q_v/k$ will be treated as being of order $k^0$. It
should be noted that upon adopting this convention all contributions from
the generalized thermodynamic forces in (\ref{2.15}) drop out from
(\ref{3.1}), as these become of higher than second order in the wave
number.

The Fourier transform of the charge conservation law (\ref{2.2}) is derived 
in a similar way. Dividing the conservation law by $k$ so as to obtain the
time derivative of $\delta q_v/k$, we see that the terms with the
generalized thermodynamic forces in (\ref{2.15}) contribute in second order
of the wave number. We get
\begin{eqnarray}
\label{3.3}
\fl \frac{\partial}{\partial t} \; \left(\frac{\delta q_v}{k}\right)=
-\rmi\, \omega_p \, \sqrt{m_v} \, \hat\bi{k}\cdot\bi{v}
-\frac{1}{T}\,\hat\bi{k}\cdot\bfsfL\cdot
\left(\hat\bi{k}\,\frac{\delta q_v}{k} 
+\rmi\, \gamma \,\sqrt{m_v} \,\bi{v}\wedge\hat\bi{B}\right) 
\nonumber\\
+k \, \hat\bi{k}\cdot\bfsfL_{\varepsilon}\cdot\hat\bi{k}\,
 \delta\left(\frac{1}{T}\right)
-k \, \frac{1}{T}\sum_{\sigma(\neq 1)} \, 
\hat\bi{k}\cdot\bfsfL_{\sigma}\cdot\hat\bi{k}\, 
\left(\frac{\delta\mu_{\sigma}}{m_{\sigma}}-
\frac{\delta\mu_1}{m_1}\right)_T 
\nonumber\\
+k^2 \, \frac{1}{T}\hat\bi{k}\hat\bi{k}:\bfsfL^{\rm (c)}:\hat\bi{k}
\left(\hat\bi{k} \, \frac{\delta q_v}{k}
+\rmi\, \gamma\, \sqrt{m_v} \,\bi{v}\wedge\hat\bi{B}\right)
\nonumber\\
-\rmi\, k^2 \frac{1}{\sqrt{m_v}\,T} \, 
\hat\bi{k}\hat\bi{k}:\bfsfL^{\rm (c)}_{v}:\hat\bi{k} \,\sqrt{m_v}\,\bi{v} 
\end{eqnarray}
In the first two lines new combinations of phenomenological coefficients
show up:
\begin{eqnarray}
\label{3.4}
\bfsfL&=&\sum_{\sigma(\neq 1)} \sum_{\sigma'(\neq 1)}
\left(\frac{e_{\sigma}}{m_{\sigma}}
-\frac{e_1}{m_1}\right)\,
\left(\frac{e_{\sigma'}}{m_{\sigma'}}
-\frac{e_1}{m_1}\right)\,\bfsfL_{\sigma\sigma'}\\
\label{3.5}
\bfsfL_{\varepsilon}&=&\sum_{\sigma(\neq 1)}\left(\frac{e_{\sigma}}{m_{\sigma}}
-\frac{e_1}{m_1}\right)\, \bfsfL_{\sigma\varepsilon}
\end{eqnarray}
Analogously, the terms in order $k^2$ contain combinations of generalized
phenomenological coefficients:
\begin{eqnarray}
\label{3.6}
\bfsfL^{\rm (c)}&=&\sum_{\sigma(\neq 1)} \sum_{\sigma'(\neq 1)}
\left(\frac{e_{\sigma}}{m_{\sigma}}
-\frac{e_1}{m_1}\right)\,
\left(\frac{e_{\sigma'}}{m_{\sigma'}}
-\frac{e_1}{m_1}\right)\,\bfsfL^{\rm (c)}_{\sigma\sigma'}\\
\label{3.7}
\bfsfL^{\rm (c)}_{v}&=&\sum_{\sigma(\neq 1)}\left(\frac{e_{\sigma}}{m_{\sigma}}
-\frac{e_1}{m_1}\right)\, \bfsfL^{\rm (c)}_{\sigma v}
\end{eqnarray}

The Fourier transform of the equation of motion follows upon inserting the
Fourier versions of (\ref{2.15}) and (\ref{2.19}) in (\ref{2.3}). Using
(\ref{2.5})--(\ref{2.6}), (\ref{2.16})--(\ref{2.17}) and (\ref{2.20}) we
get
\begin{eqnarray}
\label{3.8}
\fl \sqrt{m_v}\, \frac{\partial}{\partial t}\bi{v}=
-\rmi\,\omega_p\left(\hat{\bi{k}}\,\frac{\delta q_v}{k}
+\rmi\, \gamma\, \sqrt{m_v}\, \bi{v} \wedge\hat\bi{B}\right)
+\rmi\,\frac{\gamma}{T}\,\hat\bi{B}\wedge\bfsfL\cdot
\left(\hat\bi{k}\,\frac{\delta q_v}{k}
+\rmi\,\gamma\,\sqrt{m_v}\,\bi{v}\wedge\hat\bi{B}\right)
\nonumber\\
-\rmi\, k\, \omega_p\, \hat{\bi{k}}\, \frac{\delta p}{q_v}
-\rmi\,k\,\gamma\, \hat\bi{B}\wedge\bfsfL_{\varepsilon}\cdot\hat\bi{k}\,
 \delta\left(\frac{1}{T}\right)
\nonumber\\
+\rmi\,k \, \frac{\gamma}{T}\sum_{\sigma(\neq 1)} \, 
\hat\bi{B}\wedge\bfsfL_{\sigma}\cdot\hat\bi{k}\,
\left(\frac{\delta\mu_{\sigma}}{m_{\sigma}}-
\frac{\delta\mu_1}{m_1}\right)_T 
-k^2\,\frac{1}{\sqrt{m_v}}\, \hat\bi{k}\cdot\bfsfeta :\hat\bi{k}\,\bi{v} 
\nonumber\\
-\rmi\,k^2 \, \frac{\gamma}{T}\,
\hat\bi{B}\wedge\left[\hat\bi{k}\cdot\bfsfL^{\rm (c)}:\hat\bi{k}
\left(\hat\bi{k} \, \frac{\delta q_v}{k}
+\rmi\,\gamma \,\sqrt{m_v} \,\bi{v}\wedge\hat\bi{B}\right)\right]
\nonumber\\
-k^2\, \frac{\gamma}{\sqrt{m_v}\,T} \, 
\hat\bi{B}\wedge\left[\hat\bi{k}\cdot\bfsfL^{\rm (c)}_{v}:\hat\bi{k} \,
(\sqrt{m_v}\,\bi{v})\right]
\nonumber\\
-\rmi\, k^2 \, \frac{1}{\sqrt{m_v}\,T} \,
\hat\bi{k}\cdot\bfsfL^{\rm (c)}_{v'}:\hat\bi{k}
\left(\hat\bi{k} \,\frac{\delta q_{v}}{k} 
+\rmi\,\gamma \,\sqrt{m_v} \,\bi{v}\wedge\hat\bi{B}\right)
\end{eqnarray}
with the abbreviation
\begin{equation}
\label{3.9}
\bfsfL^{\rm (c)}_{v'}=\sum_{\sigma(\neq 1)}\left(\frac{e_{\sigma}}{m_{\sigma}}
-\frac{e_1}{m_1}\right)\, \bfsfL^{\rm (c)}_{v\sigma}
\end{equation}

Finally, we have to determine the Fourier transform of the energy law 
(\ref{2.4}). With the help of (\ref{2.9}) we get
\begin{eqnarray}
\label{3.10}
\fl \frac{\partial}{\partial t}\, \delta\varepsilon=
-\rmi\,k\,\omega_{p}\,\frac{h_v}{q_v}\,\sqrt{m_v}\, \hat\bi{k}\cdot\bi{v}\,
-k\,\frac{1}{T}\,
\hat\bi{k}\cdot\bar{\bfsfL}_{\varepsilon}\cdot
\left(\hat\bi{k}\,\frac{\delta q_v}{k} 
+\rmi\,\gamma \,
\sqrt{m_v} \,\bi{v}\wedge\hat\bi{B}\right)
\nonumber\\
+k^2 \, \hat\bi{k}\cdot\bar{\bfsfL}_{\varepsilon\varepsilon}\cdot\hat\bi{k}\,
 \delta\left(\frac{1}{T}\right)
-k^2 \, \frac{1}{T}\sum_{\sigma(\neq 1)} \, 
\hat\bi{k}\cdot\bar{\bfsfL}_{\varepsilon\sigma}\cdot\hat\bi{k}\,
\left(\frac{\delta\mu_{\sigma}}{m_{\sigma}}-
\frac{\delta\mu_1}{m_1}\right)_T 
\end{eqnarray}
with $h_v=\varepsilon+p$ the enthalpy per volume. The tensor
$\bar{\bfsfL}_{\varepsilon}$ is defined in terms of (\ref{2.11}) in the
same way as in (\ref{3.5}):
\begin{equation}
\label{3.11}
\bar{\bfsfL}_{\varepsilon}=\sum_{\sigma(\neq 1)}\left(\frac{e_{\sigma}}{m_{\sigma}}
-\frac{e_1}{m_1}\right)\, \bar{\bfsfL}_{\varepsilon\sigma}
\end{equation}

The conservation laws (\ref{3.1}) (for $\sigma\neq 1$), (\ref{3.3}),
(\ref{3.10}) and the balance equation (\ref{3.8}) form a complete set of
coupled linear differential equations. They govern the time dependence of
the perturbations in the partial densities, the charge density, the
hydrodynamical velocity and the energy density. In the following, we shall
determine the oscillating modes of the system for small wave numbers, by
solving the associated eigenvalue problem in consecutive orders of the wave
number.

\section{Oscillating modes in zeroth and first order of the wave number}
\setcounter{equation}{0}
\label{sec4}

In zeroth order of the wave number, the equations (\ref{3.1}) for the
perturbations of the partial densities $n_{\sigma}$ and (\ref{3.10}) for
the perturbation of the energy density $\varepsilon$ get a simple form, as
all terms at the right-hand side vanish for $k$ tending to 0. Hence, we are
left with the charge conservation law (\ref{3.3}) and the equation of
motion (\ref{3.8}). The latter can be decomposed in three equations for
independent components of the velocity by writing
\begin{equation}
\label{4.1}
\fl \bi{v}=\frac{1}{\hat{k}^2_{\parallel}}\,
(\hat\bi{k}_{\parallel}\cdot\bi{v})\,\hat\bi{k}_{\parallel}\,
+\frac{1}{\hat{k}^2_{\perp}}\,
(\hat\bi{k}_{\perp}\cdot\bi{v})\,\hat\bi{k}_{\perp}\,
+\frac{1}{\hat{k}^2_{\perp}}\,
[(\hat\bi{k}_{\perp}\wedge\hat\bi{B})\cdot\bi{v}]\,
\hat\bi{k}_{\perp}\wedge\hat\bi{B}
\end{equation}
The Fourier variable $\bi{k}$ has been written as
$\bi{k}=\bi{k}_{\parallel} +\bi{k}_{\perp}$, with
$\bi{B}\cdot\bi{k}_{\perp}=0$ and $\bi{B}
\wedge\bi{k}_{\parallel}=0$. Furthermore, we introduced the notations
$\hat\bi{k}_{\parallel}=\bi{k}_{\parallel}/k$, $\hat\bi{k}_{\perp}=
\bi{k}_{\perp}/k$, $\hat{k}_{\parallel}=|\hat\bi{k}_{\parallel}|$ and
$\hat{k}_{\perp}=|\hat\bi{k}_{\perp}|$. 

Using the above decomposition of (\ref{3.8}) in zeroth order of $k$ and
combining the resulting three equations with the zeroth-order approximation
to (\ref{3.3}) one gets four coupled equations from which the oscillating
modes $a^{(0)}_{\lambda\rho}$ and the associated frequencies
$z^{(0)}_{\lambda\rho}$, with $\lambda=\pm 1$ and $\rho=\pm 1$, can be
obtained \cite{GAQSLGS01}. The time dependence of the modes is determined
by the eigenvalue equation
\begin{equation}
\label{4.2}
\frac{\partial}{\partial t}\, a_{\lambda\rho}^{(0)} = 
-\rmi\, z_{\lambda\rho}^{(0)}\,a_{\lambda\rho}^{(0)}
\end{equation}
with frequencies $z_{\lambda\rho}^{(0)}$ that are solutions of the quartic
equation
\begin{eqnarray}
\label{4.3}
\fl \left[ z_{\lambda\rho}^{(0)\,2}-c \, z_{\lambda\rho}^{(0)} -\omega_p^2
\hat{k}^2_{\parallel}\right]\, \left[\left(z_{\lambda\rho}^{(0)}-\rmi\,
\gamma\, b'\right)^2 -\gamma^2 (\omega_p +b)^2\right]\nonumber\\
\rule{-1cm}{0cm}+z_{\lambda\rho}^{(0)}\, \hat{k}^2_{\perp}\,
\left\{\left(z_{\lambda\rho}^{(0)}-\rmi\, \gamma\, b'\right)\, \left[
b'^2-(\omega_p+b)^2\right] -2\rmi\, \gamma\, (\omega_p+b)^2\, b'\right\}=0
\end{eqnarray}
The coefficients $b$, $b'$ and $c$ are defined as 
\begin{equation}
\label{4.4}
\fl b=-\frac{\gamma}{T}\, L^t \quad , \quad 
b'=-\frac{\gamma}{T}\, L^{\perp} \quad , \quad 
c=-\frac{\rmi}{T}\, \left(\hat{k}^2_{\parallel}L^{\parallel}
+\hat{k}^2_{\perp}L^{\perp}\right)
\end{equation}
with $L^{\parallel}$, $L^{\perp}$ and $L^t$ independent parameters that
determine the anisotropic tensor $\bfsfL$ through the decomposition
\begin{equation}
\label{4.5}
\bfsfL=L^{\parallel}\,\hat\bi{B}\hat\bi{B}
+L^{\perp}\left(\bfsfU-\hat\bi{B}\hat\bi{B}\right)
+L^t\,\bfsfeps\cdot\hat\bi{B}
\end{equation}
with $\bfsfeps$ the Levi-Civita tensor. Since $b$ and $b'$ are real, while
$c$ is imaginary, the four solutions of (\ref{4.3}) come in pairs. Their
labels are chosen such that one has $z^*_{\lambda\rho}=-z_{\lambda,-\rho}$.

The four mode amplitudes can be written as \cite{GAQSLGS01}:
\begin{equation}
\label{4.6}
a_{\lambda\rho}^{(0)}=\frac{\delta q_v}{k}+
\sqrt{m_v}\,\bi{v}_{\lambda\rho}\cdot\bi{v}
\end{equation}
The vectors $\bi{v}_{\lambda\rho}$ have the form
\begin{equation}
\label{4.7}
\bi{v}_{\lambda\rho}=
v^{\parallel}_{\lambda\rho}\, \hat\bi{k}_{\parallel} 
+v^{\perp}_{\lambda\rho}\, \hat\bi{k}_{\perp}\, 
+v^{t}_{\lambda\rho}\, \hat\bi{k}_{\perp}\wedge\hat\bi{B}
\end{equation}
where the components are defined as
\begin{equation}
\label{4.8}
\fl v^{\parallel}_{\lambda\rho} =
\frac{\omega_p}{z_{\lambda\rho}^{(0)}}\quad , \quad 
v^{\perp}_{\lambda\rho} =
\frac{z_{\lambda\rho}^{(0)}(\omega_p+b)}{\Delta^{(0)}_{\lambda\rho}}\quad ,
\quad
v^{t}_{\lambda\rho}=\frac{z_{\lambda\rho}^{(0)}\, b'
-\rmi\, \gamma\, [b'^2+(\omega_p+b)^2]}{\Delta^{(0)}_{\lambda\rho}}
\end{equation}
with the abbreviation 
\begin{equation}
\label{4.9}
\Delta^{(0)}_{\lambda\rho}=\left(z_{\lambda\rho}^{(0)}-\rmi\, \gamma\, b'\right)^2
-\gamma^2 (\omega_p +b)^2
\end{equation}

To determine the first-order corrections to the above oscillating modes we
write
\begin{equation}
\label{4.10}
a_{\lambda\rho}^{(1)}=a_{\lambda\rho}^{(0)}\,+\,k\,a_{\lambda\rho}^{[1]}
\end{equation}
Here and in the following, a superscript $\mbox{}^{(n)}$ in round brackets
denotes the sum of all contributions up to and including a given order $n$
in $k$. A superscript $\mbox{}^{[n]}$ in square brackets denotes the
coefficient of the contribution of a specific order $n$. In particular, one has
$a_{\lambda\rho}^{(0)}=a_{\lambda\rho}^{[0]}$. As an {\em Ansatz} we will
try and solve a first-order eigenvalue equation of the form
\begin{equation}
\label{4.11}
\frac{\partial}{\partial t}\, a_{\lambda\rho}^{(1)} = \,
-\rmi\, z_{\lambda\rho}^{(0)}\,a_{\lambda\rho}^{(1)}
\end{equation}
with the same zeroth-order frequency as in (\ref{4.2}), and with a
first-order term of the general form
\begin{equation}
\label{4.12}
a_{\lambda\rho}^{[1]}= A_{\lambda\rho,\varepsilon}\, \delta\varepsilon\,
+ \sum_{\sigma(\neq 1)} A_{\lambda\rho,\sigma}\, \delta n_{\sigma}
\end{equation}
where the coefficients have to be determined as yet. Inserting
(\ref{4.10}) in (\ref{4.11}) we find 
\begin{equation}
\label{4.13}
\fl
\left[\frac{\partial}{\partial t}\, a_{\lambda\rho}^{(0)}\right]^{(0)} +
k \,\left[\frac{\partial}{\partial t}\, a_{\lambda\rho}^{(0)}\right]^{[1]} +
k\,\left[\frac{\partial}{\partial t}\, a_{\lambda\rho}^{[1]}\right]^{(0)} = 
-\rmi\, z_{\lambda\rho}^{(0)}\,a_{\lambda\rho}^{(0)}
-\rmi\, k\,z_{\lambda\rho}^{(0)}\,a_{\lambda\rho}^{[1]}
\end{equation}
up to first order in the wave number. Of course, the zeroth-order terms
drop out in view of (\ref{4.2}). Upon substitution of (\ref{4.12}) and
inspection of (\ref{3.1}) and (\ref{3.10}) one infers that the last term at
the left-hand side vanishes. Hence, we are left with the relation
\begin{equation}
\label{4.14}
\left[\frac{\partial}{\partial t}\, a_{\lambda\rho}^{(0)}\right]^{[1]} =
-\rmi\,z_{\lambda\rho}^{(0)}\,a_{\lambda\rho}^{[1]}
\end{equation}
from which the coefficients $A_{\lambda\rho,\varepsilon}$ and
$A_{\lambda\rho,\sigma}$ have to be determined.

To evaluate the left-hand side of (\ref{4.14}) we need the conservation law
for the charge density and the balance equation for the three components of
the velocity up to first order in the wave number. These follow from
(\ref{3.3}) and (\ref{3.8}), with (\ref{4.1}) inserted. To obtain more
suitable forms for these equations we rewrite the chemical potentials
$\mu_{\sigma}$ in (\ref{3.3}) and (\ref{3.8}) in terms of new chemical
potentials $\tilde{\mu}_{\sigma} =\mu_{\sigma}-(e_{\sigma}/e_1)\mu_1$
\cite{LGSAJVW87}. These are better suited to the description of an ionic
mixture, which satisfies the constraint of overall charge neutrality. The
resulting expressions for the perturbations of the chemical potentials are
given in (\ref{A2}). Furthermore, we decompose the tensors
$\bar{\bfsfL}_{\varepsilon}$ and $\bfsfL_{\sigma}$ in a similar way as in
(\ref{4.5}). In this way we arrive at the following forms of the balance
equations:
\begin{eqnarray}
\label{4.15}
\fl \frac{\partial}{\partial t} \; \left(\frac{\delta q_v}{k}\right)=
-\rmi\, c\,\frac{\delta q_v}{k} 
-\rmi\, \omega_p \, \sqrt{m_v} \, \hat\bi{k}_{\parallel}\cdot\bi{v}
-\rmi\left(\omega_p+b\right)\,  \sqrt{m_v} \, \hat\bi{k}_{\perp}\cdot\bi{v}
\nonumber\\
-\rmi\, b' \, \sqrt{m_v} \, (\hat\bi{k}_{\perp}\wedge\hat\bi{B})\cdot\bi{v}
\nonumber\\
+\rmi\, k\, \left[ T \, c_{\varepsilon}\, \delta\left(\frac{1}{T}\right) 
-T\, \sum_{\sigma(\neq 1)} c_{\sigma}\, 
\delta\left(\frac{\tilde{\mu}_{\sigma}}{T}\right)
-c\, \frac{\delta p}{q_v}\, \right]
\end{eqnarray}
\begin{equation}
\label{4.16}
\fl \sqrt{m_v}\, \frac{\partial}{\partial t}
\left(\hat\bi{k}_{\parallel}\cdot \bi{v}\right)=
-\rmi\, \omega_p\, \hat{k}_{\parallel}^2\, \frac{\delta q_v}{k}
-\rmi\, k\,  \omega_p\, \hat{k}_{\parallel}^2\,  \frac{\delta p}{q_v}
\end{equation}
\begin{eqnarray}
\label{4.17}
\fl \sqrt{m_v}\, \frac{\partial}{\partial t}
\left(\hat\bi{k}_{\perp}\cdot \bi{v}\right)=
-\rmi\,  \left(\omega_p+b\right)\, \hat{k}_{\perp}^2\, \frac{\delta q_v}{k}
+\gamma \, b'\, \sqrt{m_v}\, \hat\bi{k}_{\perp}\cdot\bi{v} \nonumber\\
-\gamma \,\left(\omega_p+ b\right)\, \sqrt{m_v}\, 
(\hat\bi{k}_{\perp}\wedge\hat\bi{B})\cdot\bi{v} \nonumber\\
+\rmi\, k\, \hat{k}^2_{\perp}\, \left[
T \, b_{\varepsilon}\, \delta\left(\frac{1}{T}\right) 
-T\, \sum_{\sigma(\neq 1)} b_{\sigma}\, 
\delta\left(\frac{\tilde{\mu}_{\sigma}}{T}\right)
-(\omega_p +b)\, \frac{\delta p}{q_v}\, \right]
\end{eqnarray}
\begin{eqnarray}
\label{4.18}
\fl \sqrt{m_v}\, \frac{\partial}{\partial t}
\left[\left(\hat\bi{k}_{\perp}\wedge\hat\bi{B}\right)\cdot \bi{v}\right]=
\rmi\, b'\,\hat{k}_{\perp}^2\,\frac{\delta q_v}{k}
+\gamma\, \left(\omega_p+ b\right)\,
\sqrt{m_v}\, \hat\bi{k}_{\perp}\cdot\bi{v} \nonumber\\
+\gamma \, b'\, \sqrt{m_v}\, 
(\hat\bi{k}_{\perp}\wedge\hat\bi{B})\cdot \bi{v} \nonumber\\
-\rmi\, k\, \hat{k}^2_{\perp}\, \left[
T \, b'_{\varepsilon}\, \delta\left(\frac{1}{T}\right) 
-T\, \sum_{\sigma(\neq 1)} b'_{\sigma}\, 
\delta\left(\frac{\tilde{\mu}_{\sigma}}{T}\right)
-b'\, \frac{\delta p}{q_v}\, \right]
\end{eqnarray}
The abbreviations $b_{\varepsilon}$, $b'_{\varepsilon}$, $b_{\sigma}$,
$b'_{\sigma}$, $c_{\varepsilon}$ and $c_{\sigma}$ are combinations of the
components of $\bfsfL$, $\bar{\bfsfL}_{\varepsilon}$ and $\bfsfL_{\sigma}$
\cite{GAQSLGS01}:
\begin{equation}
\label{4.19}
b_{\varepsilon}=-\frac{\gamma}{T}\, \bar{L}^t_{\varepsilon}
-b\, \frac{h_v}{q_v} \quad , \quad 
b'_{\varepsilon}=-\frac{\gamma}{T}\, \bar{L}^{\perp}_{\varepsilon}
-b'\, \frac{h_v}{q_v} 
\end{equation}
\begin{equation}
\label{4.20}
b_{\sigma}=-\frac{\gamma}{m_{\sigma}T}\, L^t_{\sigma}
-b\, \frac{n_{\sigma}}{q_v} \quad , \quad 
b'_{\sigma}=-\frac{\gamma}{m_{\sigma}T}\, L^{\perp}_{\sigma}
-b'\, \frac{n_{\sigma}}{q_v} 
\end{equation}
\begin{equation}
\label{4.21}
\fl c_{\varepsilon}=-\frac{\rmi}{T}\, 
\left(\hat{k}^2_{\parallel}\bar{L}^{\parallel}_{\varepsilon}
+\hat{k}^2_{\perp}\bar{L}^{\perp}_{\varepsilon}\right)-c\, \frac{h_v}{q_v} 
\quad , \quad 
c_{\sigma}=-\frac{\rmi}{m_{\sigma}T}\, 
\left(\hat{k}^2_{\parallel}L^{\parallel}_{\sigma}
+\hat{k}^2_{\perp}L^{\perp}_{\sigma}\right)-c\, \frac{n_{\sigma}}{q_v} 
\end{equation}
with $\sigma\neq 1$.

The terms of order $k$ in (\ref{4.15})--(\ref{4.18}) contain the
perturbations $\delta(1/T)$, $\delta(\tilde{\mu}_{\sigma}/T)$ and $\delta
p$. These can be expressed in terms of the perturbations $\delta
n_{\sigma}$ and $\delta\varepsilon$ by employing the thermodynamic
relations (\ref{A9}), (\ref{A10}) and (\ref{A13}) of \ref{appA} of which
only the terms of order $k^0$ are needed for the moment. Substituting the
results in (\ref{4.14}) with (\ref{4.6}), we arrive at an
expression for $a^{[1]}_{\lambda\rho}$, which is indeed of the form of
(\ref{4.12}). Upon using (\ref{4.3}) we obtain the following expression for
the coefficients:
\begin{eqnarray}
\label{4.22}
A_{\lambda\rho,j} = \beta^{-1}\left( \frac{n}{2\beta q_v} \,
M^{-1}_{\varepsilon j} - s_j + \sum_{j'} M^{-1}_{jj'}\,F_{\lambda\rho,j'}\right)
\end{eqnarray}
with $j$ and $j'$ equal to $\varepsilon$ or $\sigma (\neq 1)$, with
$\beta=1/(k_BT)$, and with $M^{-1}_{jj'}$ and $s_j$ defined in (\ref{A4})
and (\ref{A8}). Furthermore, we introduced the abbreviation
\begin{equation}
\label{4.23}
F_{\lambda\rho,j}=
\frac{1}{z_{\lambda\rho}^{(0)}}\, \left[ c_j +
\left(v^{\perp}_{\lambda\rho}\, b_j-
v^{t}_{\lambda\rho}\, b'_j\right)\hat{k}_{\perp}^2\right]
\end{equation}
with components $v^i_{\lambda\rho}$ defined in (\ref{4.8}).

\section{Oscillating-mode frequencies in second order of the wave number}
\setcounter{equation}{0}
\label{sec5}
Having determined the oscillating modes up to first order in the wave
number, we now turn to the contributions of second order in $k$. In
particular, we want to evaluate the mode frequencies in second order, as
these determine the dispersion and damping effects. The mode amplitudes in
second order are less interesting. Fortunately, their explicit form is not
needed in the following, as we shall see.

The second-order contributions $z^{[2]}_{\lambda\rho}$ to the
eigenfrequencies of the oscillating modes follow from the eigenvalue
equation
\begin{equation}
\label{5.1}
\frac{\partial}{\partial t}\, a_{\lambda\rho}^{(2)} = 
-\rmi\, z_{\lambda\rho}^{(2)}\,a_{\lambda\rho}^{(2)}
\end{equation}
with the mode frequencies 
\begin{equation}
\label{5.2}
z_{\lambda\rho}^{(2)} = z_{\lambda\rho}^{(0)} + k^2\,z_{\lambda\rho}^{[2]}
\end{equation}
and the mode amplitudes
\begin{equation}
\label{5.3}
a_{\lambda\rho}^{(2)}=a_{\lambda\rho}^{(0)}
+ k\,a_{\lambda\rho}^{[1]} + k^2\,a_{\lambda\rho}^{[2]}
\end{equation}
The zeroth- and first-order terms in the amplitudes have been given in
(\ref{4.6}) and (\ref{4.12}) with (\ref{4.22}). The second-order terms have
the general form
\begin{equation}
\label{5.4}
\fl a_{\lambda\rho}^{[2]} = A_{\lambda\rho,q} \, \frac{\delta q_v}{k}
+A_{\lambda\rho}^{\parallel}\, \sqrt{m_v} \, \hat\bi{k}_{\parallel}\cdot\bi{v}
+A_{\lambda\rho}^{\perp}\, \sqrt{m_v}\, \hat\bi{k}_{\perp}\cdot\bi{v}
+A_{\lambda\rho}^{t}\, \sqrt{m_v}\, (\hat\bi{k}_{\perp}\wedge\hat\bi{B})\cdot\bi{v}
\end{equation}
with coefficients $A_{\lambda\rho}$ that are as yet unknown.

Substitution of (\ref{5.2}) and (\ref{5.3}) into (\ref{5.1}) yields the
second-order equation
\begin{equation}
\label{5.5}
\fl
\left[\frac{\partial}{\partial t}\, a_{\lambda\rho}^{(0)}\right]^{[2]} + 
\left[\frac{\partial}{\partial t}\, a_{\lambda\rho}^{[1]}\right]^{[1]} + 
\left[\frac{\partial}{\partial t}\, a_{\lambda\rho}^{[2]}\right]^{(0)} = \,
-\rmi\, z_{\lambda\rho}^{(0)}\,a_{\lambda\rho}^{[2]}-\rmi\,
z_{\lambda\rho}^{[2]}\,a_{\lambda\rho}^{(0)}
\end{equation}
The right-hand side is a linear combination of the four independent
perturbations $\delta q_v/k$, $\sqrt{m_v}\,
\hat\bi{k}_{\parallel}\cdot\bi{v}$, $\sqrt{m_v}\,
\hat\bi{k}_{\perp}\cdot\bi{v}$ and $\sqrt{m_v}\,
(\hat\bi{k}_{\perp}\wedge\hat\bi{B})\cdot\bi{v}$, as follows by inserting
(\ref{4.6})--(\ref{4.7}) and (\ref{5.4}). Likewise, the third term at the
left-hand side of (\ref{5.5}) can be written in terms of these independent
perturbations by inserting (\ref{5.4}) and substituting the zeroth-order
contributions of (\ref{4.15})--(\ref{4.18}). In the second term at the
left-hand side of (\ref{5.5}) we substitute (\ref{4.12}) and use
(\ref{3.1}), (\ref{3.10}). The first-order contributions in the latter
equations can be rewritten as linear combinations of the independent
perturbations in the way shown in (6.1)--(6.2) of
\cite{GAQSLGS01}. Finally, we have to consider the first term at the
left-hand side of (\ref{5.5}). Formally, it can be written as a linear
combination of the independent perturbations as well:
\begin{equation}
\label{5.6}
\fl \left[\frac{\partial}{\partial t}\, a_{\lambda\rho}^{(0)}\right]^{[2]}=
D_{\lambda\rho,q} \, \frac{\delta q_v}{k}
+D_{\lambda\rho}^{\parallel}\, \sqrt{m_v} \, \hat\bi{k}_{\parallel}\cdot\bi{v}
+D_{\lambda\rho}^{\perp}\, \sqrt{m_v}\, \hat\bi{k}_{\perp}\cdot\bi{v}
+D_{\lambda\rho}^{t}\, \sqrt{m_v}\, (\hat\bi{k}_{\perp}\wedge\hat\bi{B})\cdot\bi{v}
\end{equation}
The coefficients $D_{\lambda\rho}$ will be determined later.

Having taken the above steps we are in a position to compare the
contributions of the four independent perturbations at both sides of
(\ref{5.5}). Upon equating the coefficients we arrive at the following set
of equalities:
\begin{eqnarray}
\label{5.7}
\fl 
(c-z_{\lambda\rho}^{(0)})\, A_{\lambda\rho,q}
+\omega_p\,\hat{k}_{\parallel}^2\,A_{\lambda\rho}^{\parallel}
+(\omega_p+b)\,\hat{k}_{\perp}^2\,A_{\lambda\rho}^{\perp}
-b'\,\hat{k}_{\perp}^2\,A_{\lambda\rho}^{t}=
z_{\lambda\rho}^{[2]}-\rmi \, D_{\lambda\rho,q}
\nonumber\\
-A_{\lambda\rho,\varepsilon}\left(c_{\varepsilon}+c\,
\frac{h_v}{q_v}\right)
-\sum_{\sigma(\neq 1)}A_{\lambda\rho,\sigma}\left(c_{\sigma}+c\,
\frac{n_{\sigma}}{q_v}\right)
\end{eqnarray}
\begin{eqnarray}
\label{5.8}
\fl
\omega_p\,A_{\lambda\rho,q}-
z_{\lambda\rho}^{(0)}\,A_{\lambda\rho}^{\parallel}
=z_{\lambda\rho}^{[2]}\, v^{\parallel}_{\lambda\rho}
-\rmi \, D_{\lambda\rho}^{\parallel}
-A_{\lambda\rho,\varepsilon}\,\omega_p\,\frac{h_v}{q_v}
-\sum_{\sigma(\neq 1)}A_{\lambda\rho,\sigma}\,\omega_p\,\frac{n_{\sigma}}{q_v}
\end{eqnarray}
\begin{eqnarray}
\label{5.9}
\fl
(\omega_p+b)\, A_{\lambda\rho,q}
+(\rmi\,\gamma\, b'-z_{\lambda\rho}^{(0)})\,A_{\lambda\rho}^{\perp}
+\rmi\,\gamma\, (\omega_p+b)\, A_{\lambda\rho}^{t}=
z_{\lambda\rho}^{[2]}\,v^{\perp}_{\lambda\rho}
-\rmi \, D_{\lambda\rho}^{\perp}\nonumber\\
-A_{\lambda\rho,\varepsilon}\left[b_{\varepsilon}+(\omega_p+b)\,
\frac{h_v}{q_v}\right]
-\sum_{\sigma(\neq 1)}A_{\lambda\rho,\sigma}\left[b_{\sigma}+(\omega_p+b)\,
\frac{n_{\sigma}}{q_v}\right]
\end{eqnarray}
\begin{eqnarray}
\label{5.10}
\fl
b'\,A_{\lambda\rho,q}-\rmi\,\gamma\,(\omega_p+b)\,A_{\lambda\rho}^{\perp}
+(\rmi\,\gamma\, b'-z_{\lambda\rho}^{(0)})\, A_{\lambda\rho}^{t}=
z_{\lambda\rho}^{[2]}\, v^{t}_{\lambda\rho}
-\rmi \, D_{\lambda\rho}^{t}\nonumber\\
-A_{\lambda\rho,\varepsilon}\left(b'_{\varepsilon}+b'\,
\frac{h_v}{q_v}\right)
-\sum_{\sigma(\neq 1)}A_{\lambda\rho,\sigma}\left(b'_{\sigma}+b'\,
\frac{n_{\sigma}}{q_v}\right)
\end{eqnarray}

The above system of equations for the four unknowns $A_{\lambda\rho,q}$,
$A_{\lambda\rho}^{\parallel}$, $A_{\lambda\rho}^{\perp}$ and
$A_{\lambda\rho}^{t}$ turns out to be a dependent set. In fact, the
determinant of the matrix of coefficients appearing at the left-hand sides
is exactly equal to the left-hand side of the quartic equation (\ref{4.3}),
so that it vanishes. The mutual dependence of the equations is a
mathematical reflection of the fact that the amplitudes of the modes are
fixed up to a multiplicative constant only. Because the system is a dependent
set, the matrix has got a left-eigenvector with eigenvalue 0. This
four-dimensional vector is found as
\begin{equation}
\label{5.11}
\left(1\, ,\, 
v^{\parallel}_{\lambda\rho} \,\hat{k}_{\parallel}^2 \, ,\,
v^{\perp}_{\lambda\rho} \,\hat{k}_{\perp}^2\, ,\,
-v^{t}_{\lambda\rho}\,\hat{k}_{\perp}^2\right)
\end{equation}
The second-order contributions $z_{\lambda\rho}^{[2]}$ to the
eigenfrequencies of the oscillating modes now follow by taking
the inner product of (\ref{5.11}) with the set of equation
(\ref{5.7})--(\ref{5.10}). The left-hand side of the resulting equation
vanishes by construction, so that we are left with an equality from which
$z_{\lambda\rho}^{[2]}$ can be determined. Using (\ref{4.3}) to eliminate
$c$ we find:
\begin{eqnarray}
\label{5.12}
\fl z^{[2]}_{\lambda\rho}=N_{\lambda\rho}\,\left\{
z^{(0)}_{\lambda\rho}\,\left[ 
A_{\lambda\rho,\varepsilon}\,
\left(F_{\lambda\rho,\varepsilon}+\frac{h_v}{q_v}\right)+
\sum_{\sigma(\neq 1)}A_{\lambda\rho,\sigma}\,
\left(F_{\lambda\rho,\sigma}+\frac{n_{\sigma}}{q_v}\right)\right]\right.
\nonumber\\
\left.+\rmi\, D_{\lambda\rho,q}+\rmi\,
v^{\parallel}_{\lambda\rho}\,
\hat{k}^2_{\parallel}\, D_{\lambda\rho}^{\parallel}
+\rmi\, v^{\perp}_{\lambda\rho}\,
\hat{k}_{\perp}^2\, D_{\lambda\rho}^{\perp}
-\rmi\, 
v^{t}_{\lambda\rho}\,
\hat{k}_{\perp}^2\, D_{\lambda\rho}^t\rule{0cm}{0.6cm}\right\}
\end{eqnarray}
where the constant $N_{\lambda\rho}$ is defined as
\begin{equation}
\label{5.13}
N_{\lambda\rho}=\left[1+\left(v^{\parallel}_{\lambda\rho}\right)^2
\hat{k}_{\parallel}^2
+\left(v^{\perp}_{\lambda\rho}\right)^2\,
\hat{k}_{\perp}^2
-\left(v^{t}_{\lambda\rho}\right)^2\,
\hat{k}_{\perp}^2\right]^{-1}
\end{equation}

The expression (\ref{5.12}) still contains the coefficients
$D_{\lambda\rho}$ that have been introduced in (\ref{5.6}). These will now
be determined by substituting (\ref{4.6}) in the left-hand side of
(\ref{5.6}). Employing the balance equations
(\ref{4.15})--(\ref{4.18}) and using the thermodynamical relations of
\ref{appA}, we encounter contributions that follow from the terms
proportional to $\delta q_v$ in (\ref{A9}), (\ref{A10}) and
(\ref{A13}). The ensuing term in $D_{\lambda\rho,q}$ is:
\begin{equation}
\label{5.14}
D'_{\lambda\rho,q}=\frac{\rmi}{\beta}\,
z^{(0)}_{\lambda\rho}\left(\sum_j F_{\lambda\rho,j}\,W_j+\frac{n}{2\beta q_v}\,
W_{\varepsilon} -\frac{1}{2q_ve_1}\right)
\end{equation}
A further contribution $D''_{\lambda\rho,q}$ to $D_{\lambda\rho,q}$ follows
by considering the terms of order $k^2$ in (\ref{3.3}) and
(\ref{3.8}). These yield:
\begin{eqnarray}
\label{5.15}
\fl
D''_{\lambda\rho,q}=-\frac{\rmi}{\sqrt{m_v}\,T}
\,\bi{v}_{\lambda\rho}\hat\bi{k}:\bfsfL^{{\rm (c)}}_{v'}:
\hat\bi{k}\hat\bi{k}
+\frac{1}{T}
\left(\hat\bi{k}-\rmi\,\gamma\,\bi{v}_{\lambda\rho}\wedge\hat\bi{B}\right)
\hat\bi{k}:\bfsfL^{\rm (c)}:\hat\bi{k}\hat\bi{k}
\end{eqnarray}
Likewise, we find for the other coefficients:
\begin{eqnarray}
\label{5.16}
\fl
D^{\parallel}_{\lambda\rho}= 
-\frac{1}{m_v\,\hat{k}_{\parallel}^2}\, \bi{v}_{\lambda\rho}\hat\bi{k}:\bfsfeta
:\hat\bi{k}\hat\bi{k}_{\parallel} 
- \frac{\rmi}{\sqrt{m_v}\,T\hat{k}_{\parallel}^2} \,
\left(\hat\bi{k}-\rmi\,\gamma\,\bi{v_{\lambda\rho}}\wedge\hat\bi{B}\right)
\hat\bi{k}:\bfsfL^{{\rm (c)}}_{v}:\hat\bi{k}\hat\bi{k}_{\parallel}
\end{eqnarray}
\begin{eqnarray}
\label{5.17}
\fl
D^{\perp}_{\lambda\rho}=
    -\frac{1}{m_v\,\hat{k}_{\perp}^2}\, \bi{v}_{\lambda\rho}\hat\bi{k}:\bfsfeta
:\hat\bi{k}\hat\bi{k}_{\perp} \,
+\frac{\rmi\,\gamma}{T\hat{k}_{\perp}^2}\,
\left(\hat\bi{k}-\rmi\,\gamma\,\bi{v}_{\lambda\rho}\wedge\hat\bi{B}\right)
\hat\bi{k}:\bfsfL^{{\rm (c)}}:\hat\bi{k}(\hat\bi{k}_{\perp}\wedge\hat\bi{B})
\nonumber\\
-\frac{\rmi}{\sqrt{m_v}\,T\hat{k}_{\perp}^2} \,
\left(\hat\bi{k}-\rmi\,\gamma\,\bi{v}_{\lambda\rho}\wedge\hat\bi{B}\right)
\hat\bi{k}:\bfsfL^{{\rm (c)}}_{v}:\hat\bi{k} \hat\bi{k}_{\perp}
\nonumber\\
+\frac{\gamma}{\sqrt{m_v}\,T\hat{k}_{\perp}^2} \,
\bi{v}_{\lambda\rho}\hat\bi{k}:\bfsfL^{{\rm (c)}}_{v'}:\hat\bi{k} 
(\hat\bi{k}_{\perp}\wedge\hat\bi{B})
\end{eqnarray}
\begin{eqnarray}
\label{5.18}
\fl
D^{t}_{\lambda\rho}=
-\frac{1}{m_v\,\hat{k}_{\perp}^2}\, 
\bi{v}_{\lambda\rho}\hat\bi{k}:\bfsfeta:
\hat\bi{k}(\hat\bi{k}_{\perp}\wedge\bi{B})
-\frac{\rmi\,\gamma}{T\hat{k}_{\perp}^2} \,
\left(\hat\bi{k}-\rmi\,\gamma\,\bi{v}_{\lambda\rho}\wedge\bi{B}\right)
\hat\bi{k}:\bfsfL^{{\rm (c)}}:\hat\bi{k} \hat\bi{k}_{\perp} \,
\nonumber\\
-\frac{\rmi}{\sqrt{m_v}\,T\hat{k}_{\perp}^2} \,
\left(\hat\bi{k}-\rmi\,\gamma\,\bi{v}_{\lambda\rho}\wedge\hat\bi{B}\right)\
\hat\bi{k}:\bfsfL^{{\rm (c)}}_{v}:\hat\bi{k}  
(\hat\bi{k}_{\perp}\wedge\hat\bi{B})
\nonumber\\ 
-\frac{\gamma}{\sqrt{m_v}\,T\hat{k}_{\perp}^2} \, 
\bi{v}_{\lambda\rho}\hat\bi{k}:\bfsfL^{{\rm (c)}}
_{v'}:\hat\bi{k} \hat\bi{k}_{\perp}
\end{eqnarray}
The fourth-rank tensors in these expressions may be decomposed in terms of
their invariant parts in a way analogous to (\ref{4.5}). The
expressions for the coefficients $D_{\lambda\rho}$ that result by
substituting these decompositions are rather unwieldy and will not be
presented here.

Apart from the coefficients $D_{\lambda\rho}$, the expression (\ref{5.12})
also depends on the coefficients $A_{\lambda\rho}$ for which the explicit
form (\ref{4.22}) is available. Upon inserting the latter and combining the
resulting contributions with those from (\ref{5.14}) we arrive at the final
form for the second-order terms in the eigenfrequencies of the
oscillating modes:
\begin{eqnarray}
\label{5.19}
\fl z^{[2]}_{\lambda\rho}=N_{\lambda\rho}\,\left\{
z^{(0)}_{\lambda\rho}\,\left[ 
\frac{c_s^2}{\omega_p^2}
+\frac{2}{\beta}\sum_j\left(\frac{n}{2\beta q_v}\, M^{-1}_{\varepsilon
  j}-s_j\right) F_{\lambda\rho,j}
+\frac{1}{\beta}\sum_{j,j'} F_{\lambda\rho,j}\, M^{-1}_{jj'}\,
F_{\lambda\rho,j'}
\right]
\right.\nonumber\\
\left. +\rmi\, D''_{\lambda\rho,q}+\rmi\,v^{\parallel}_{\lambda\rho}\,
\hat{k}^2_{\parallel}\, D_{\lambda\rho}^{\parallel}
+\rmi\,v^{\perp}_{\lambda\rho}\,
\hat{k}_{\perp}^2\, D_{\lambda\rho}^{\perp}
-\rmi\, v^{t}_{\lambda\rho}\,
\hat{k}_{\perp}^2\, D_{\lambda\rho}^t\rule{0cm}{0.6cm}\right\}
\end{eqnarray}
where we have used (\ref{A18}) as well. Our result shows that the damping
and the dispersion of the oscillating modes in second order of the wave
number is caused by several rather different mechanisms. The first term,
with the sound velocity $c_s$, depends on equilibrium properties only. Of
course, dispersion of collective modes that is governed by the sound
velocity is a well-known effect, both in plasmas and in systems of neutral
particles. The remaining terms in the first line are linear or quadratic in
$F_{\lambda\rho}$. According to (\ref{4.23}) these coefficients are linear
combinations of $b_j$, $b'_j$ and $c_j$, which have been defined in
(\ref{4.19})--(\ref{4.21}) with (\ref{4.4}). They are governed by
non-equilibrium processes through the components of the phenomenological
tensors $\bfsfL$, $\bar{\bfsfL}_{\varepsilon}$ and $\bfsfL_{\sigma}$, which
are linear combinations of $\bfsfL_{\sigma\sigma'}$ and
$\bar{\bfsfL}_{\varepsilon\sigma}$. The latter occur in the linear laws
(\ref{2.7}) and (\ref{2.9}) for the diffusion flows and the heat
flow. Finally, the second line of (\ref{5.19}) depends on non-equilibrium
processes as well via the coefficients $D_{\lambda\rho}$. From
(\ref{5.15})--(\ref{5.18}) we infer that viscous effects play a role
here. However, these are not the only dissipative phenomena that are
relevant for the damping and the dispersion. Higher-order couplings between
thermodynamic forces and flows, as described by Burnett terms, are
important as well.

For the one-component plasma the expression (\ref{5.19}) gets a simpler
form. For that system $F_{\lambda\rho,j}$ vanishes, so that in the first
line only the term with the sound velocity survives. Furthermore, the
Burnett contributions disappear from the coefficients $D_{\lambda\rho}$, so
that only the viscous terms in (\ref{5.16})--(\ref{5.18}) remain. The
resulting expression agrees with that found in \cite{LGSJSC85} and
\cite{LGSAJS87a}. 

\section{Relation to kinetic theory}
\label{sec6}
\setcounter{equation}{0}

The collective modes of an ionic mixture can also be derived with the help
of formal kinetic theory \cite{AJSLGS90,GAQSLGS01}. In that approach the starting
point is the set of microscopic balance equations. From these one obtains 
the collective modes by employing a projection-operator technique. 

The microscopic partial particle density is $n_{\sigma}^{\rm
m}(\bi{r})=\sum_{\alpha}\delta (\bi{r}- \bi{r}_{\sigma\alpha})$, with $
\bi{r}_{\sigma \alpha}$ the position of particle $\alpha$ of component
$\sigma$.  The Fourier transform $\delta n_{\sigma}^{\rm m}(\bi{k})$ of the
perturbation $n_{\sigma}^{\rm m}(\bi{r})- \langle n_{\sigma}^{\rm
m}(\bi{r})\rangle$ of the microscopic partial particle density , with
$\langle n_{\sigma}^{\rm m}(\bi{r})\rangle$ its uniform thermal equilibrium
average, reads
\begin{equation}
\label{6.1}
\delta n_{\sigma}^{\rm m}(\bi{k}) = \sum_{\alpha} \rme^{- \rmi \bi{k} \cdot
\bi{r}_{\sigma \alpha}}                                        
\end{equation}
The Fourier transform of the perturbation of the microscopic charge 
density is $\delta q_v^{\rm m}(\bi{k}) = \sum_{\sigma} e_{\sigma} \delta 
n_{\sigma}^{\rm m}(\bi{k})$. Furthermore, the microscopic partial momentum 
density of species $\sigma$ in Fourier space is given by
\begin{equation}
\label{6.2}
\bi{g}_{\sigma}^{\rm m}(\bi{k}) = \sum_{\alpha} \bi{p}_{\sigma \alpha}
\rme ^{- \rmi \bi{k} \cdot \bi{r}_{\sigma \alpha}} 
\end{equation}
with $\bi{p}_{\sigma \alpha}$ the momentum of particle $\alpha$ of component 
$\sigma$. Finally, the perturbation of the microscopic 
energy density in Fourier space will be denoted by $\delta\varepsilon^{\rm 
m}(\bi{k})$. 

The conservation laws for $\delta n_{\sigma}^{\rm m}(\bi{k})$ and $\delta 
\varepsilon^{\rm m}(\bi{k})$, and the balance equation for the microscopic 
total momentum density $\bi{g}^{\rm m}(\bi{k}) = \sum_{\sigma} 
\bi{g}_{\sigma}^{\rm m}(\bi{k})$ read \cite{AJSLGS90}:
\begin{eqnarray}
\label{6.3}
\calL\, \delta n_{\sigma}^{\rm m}(\bi{k}) 
=- \frac{\bi{k}}{m_{\sigma}} \cdot \bi{g}_{\sigma}^{\rm m}(\bi{k})\\
\label{6.4}
\calL \, \bi{g}^{\rm m}(\bi{k}) 
= - \bi{k} \cdot \bi{\delta \bfsftau}^{\rm m}(\bi{k}) -q_v 
\frac{\bi{k}}{k^2} \, \delta q_v^{\rm m}(\bi{k}) 
- \rmi \sum_{\sigma} \frac{e_{\sigma}}{m_{\sigma}c}
\bi{g}_{\sigma}^{\rm m}(\bi{k}) \wedge \bi{B} \\
\label{6.5}
\calL\, \delta \varepsilon^{\rm m}(\bi{k}) 
= - \bi{k} \cdot \bi{j}_{\varepsilon}^{\rm m}(\bi{k})
\end{eqnarray}
with $\delta \bfsftau^{\rm m}(\bi{k})$ the Fourier-transformed perturbation
of the microscopic pressure tensor and $\bi{j}_{\varepsilon}^{\rm
m}(\bi{k})$ the microscopic energy-current density in Fourier
space. The Liouville operator in phase space $\calL$ follows by writing the
time derivative $\dot{F}$ of an arbitrary function $F$ as $\dot{F} = \rmi
\calL F$. 

The collective modes are specific independent linear combinations of the
microscopic basic quantities
\begin{equation}
\label{6.6}
 a_i^{\rm b}(\bi{k}) \in \left\{ \frac{1}{k} \delta q_v^{\rm m}(\bi{k}) , 
\bi{g}^{\rm m}(\bi{k}) , \delta \varepsilon^{\rm m}(\bi{k}), 
 \delta n_2^{\rm m}(\bi{k}) , \ldots , \delta n_s^{\rm m}(\bi{k})
 \right\} 
\end{equation}
with $i=0, 1, 2, 3, \varepsilon, \sigma\, (\neq 1)$. The microscopic charge
density is divided by the wave number, since the fluctuations in the charge
density vanish in the long-wavelength limit, as has been remarked already
in Section \ref{sec3}. Indeed, in leading order of the wave number one has
\begin{equation}
\label{6.7}
\frac{1}{V}\langle [\delta q_v^{\rm m}(\bi{k})]^{\ast} \delta q_v^{\rm m}
(\bi{k}) \rangle = \frac{k^2}{\beta}
\end{equation}
with $V$ the volume of the system. The fluctuation formulas for the ionic 
mixture have been given in \cite{AJSLGS90} and \cite{AJVWLGS87}.

To derive the collective modes one may use a projection-operator
technique. The one-sided Fourier transform of the time-dependent collective
mode $a^{\rm m}_i(\bi{k},t)$ is given by
\begin{equation}
\label{6.8}
a^{\rm m}_i(\bi{k},z)\equiv -\rmi\int_0^{\infty}dt\,  \rme^{\rmi zt}\,
a^{\rm m}_i(\bi{k},t)=\frac{1}{z+\calL}\, a^{\rm m}_i(\bi{k})
\end{equation}
which is regular for $z$ in the upper halfplane. By introducing a
projection operator $P$ that projects an arbitrary phase function
$f(\bi{k})$ on the space spanned by the basis (\ref{6.6}) and its
complement $Q=1-P$, one arrives at an eigenvalue problem for the modes
$a^{\rm m}_i(\bi{k})$ and their frequencies $z^{\rm m}_i(\bi{k})$:
\begin{eqnarray}
\label{6.9}
\fl -\frac{1}{V}\langle \left[a_j^{\rm b}(\bi{k})\right]^{\ast}\, 
\calL a^{\rm m}_i(\bi{k})\rangle
+\frac{1}{V}\langle \left[a_j^{\rm b}(\bi{k})\right]^{\ast}\, 
\calL Q\frac{1}{z^{\rm m}_i(\bi{k})+Q\calL
  Q} Q\calL a^{\rm m}_i(\bi{k})\rangle\nonumber\\
=z^{\rm m}_i(\bi{k})\,\frac{1}{V}\langle \left[a_j^{\rm b}(\bi{k})\right]^{\ast}\, 
a^{\rm m}_i(\bi{k})\rangle
\end{eqnarray}
for all basis functions $a_j^{\rm b}(\bi{k})$. Near $z=z^{\rm m}_i(\bi{k})$ the
projection of (\ref{6.8}) has a pole structure of the form
\begin{equation}
\label{6.10}
P\,\frac{1}{z+\calL}\, a^{\rm m}_i(\bi{k})=\frac{1}{z-z^{\rm m}_i(\bi{k})}\,
a^{\rm m}_i(\bi{k})
\end{equation}

The eigenvalue problem (\ref{6.9}) leads to a set of $s+4$ modes. Apart
from a heat mode and $s-1$ diffusion modes one finds four modes that are
the kinetic counterparts of the oscillating modes. In zeroth
order of the wave number their frequencies $z^{{\rm m}(0)}_{\lambda\rho}$ follow
from a relation of the form (\ref{4.3}). The parameters $b$, $b'$ and $c$
are replaced by $z$-dependent functions which follow from the identities
\cite{GAQSLGS01}
\begin{eqnarray}
\label{6.11}
\fl\frac{\beta}{V}
\langle \left[\frac{\delta q^{\rm m}_v(\bi{k})}{k}\right]^{\ast} \calL Q
\frac{1}{z+Q\calL Q} Q\calL \frac{\delta q^{\rm m}_v(\bi{k})}{k}\rangle^{(0)}=
c(z)\\
\label{6.12}
\fl\frac{\beta}{\sqrt{m_v}\,V}
\langle \left[\bi{g}^{\rm m}(\bi{k})\right]^{\ast} \calL Q
\frac{1}{z+Q\calL Q} Q\calL \frac{\delta q^{\rm m}_v(\bi{k})}{k}\rangle^{(0)}=
b(z)\,\hat\bi{k}_{\perp}+b'(z)\, \hat\bi{k}_{\perp}\wedge \hat\bi{B}\\
\label{6.13}
\fl\frac{\beta}{m_vV}
\langle \left[\bi{g}^{\rm m}(\bi{k})\right]^{\ast} \calL Q
\frac{1}{z+Q\calL Q} Q\calL \bi{g}^{\rm m}(\bi{k})\rangle^{(0)}
=-\rmi\gamma b(z)\,\bfsfeps\cdot\hat\bi{B}+
\rmi\gamma b'(z)\, (\bfsfU-\hat\bi{B}\hat\bi{B})
\end{eqnarray}
The consistency of the second and third of these definitions has been
demonstrated in \cite{GAQSLGS01}. 

The amplitudes $a_{\lambda\rho}^{\rm m}(\bi{k})$ of the associated modes
in zeroth order of the wave number are given by
\begin{equation} 
\label{6.14}
a_{\lambda \rho}^{{\rm m}(0)}(\bi{k}) = \frac{\delta q_v^{\rm m}(\bi{k})}{k}  +
\frac{1}{\sqrt{m_v}} \bi{v}_{\lambda \rho}^{\rm m}(\bi{k}) \cdot 
\bi{g}^{\rm m}(\bi{k})
\end{equation}
The vector $ \bi{v}_{\lambda \rho}^{\rm m}(\bi{k})$ has the same form as
(\ref{4.7})--(\ref{4.8}), with the parameters $b$, $b'$ and $c$ replaced by
the corresponding functions $b(z)$, $b'(z)$ and $c(z)$ for
$z=z^{{\rm m}(0)}_{\lambda\rho}$.

The first-order contributions to the mode amplitudes have the
general form (cf.\ (\ref{4.12}))
\begin{equation}
\label{6.15}
a^{{\rm m}[1]}_{\lambda\rho}(\bi{k})= 
A^{\rm m}_{\lambda\rho,\varepsilon}\, \delta\varepsilon^{\rm m}(\bi{k})
+ \sum_{\sigma(\neq 1)} A^{\rm m}_{\lambda\rho,\sigma}\, 
\delta n_{\sigma}^{\rm m}(\bi{k})
\end{equation}
with coefficients $A^{\rm m}_{\lambda\rho,j}$ that have to be determined by
substituting (\ref{6.14}) and (\ref{6.15}) into (\ref{6.9}) for
$i=\lambda\rho$ and $j=\varepsilon,\sigma$. Employing the fluctuation
formulas \cite{AJSLGS90}, \cite{AJVWLGS87} and the thermodynamic
formulas of \ref{appA}, we find a set of inhomogeneous linear equations for
the unknown $A^{\rm m}_{\lambda\rho,j}$, with a coefficient matrix given by
(\ref{A4}). Solving these equations we arrive at expressions for $A^{\rm
m}_{\lambda\rho,j}$ of the same form as (\ref{4.22}), with
$F_{\lambda\rho,j}$ replaced by $F^{\rm m}_{\lambda\rho,j}$, with a similar
structure as in (\ref{4.23}). The kinetic counterparts of
(\ref{4.19})--(\ref{4.21}) are \cite{GAQSLGS01}:
\begin{eqnarray}
\label{6.16}
\fl 
\frac{\beta}{V}\,\langle\left[\frac{\delta q_v^{\rm m}(\bi{k})}{k}\right]^{\ast} 
\calL Q \frac{1}{z+Q\calL Q} Q\calL\, \left[\delta\varepsilon^{\rm
      m}(\bi{k}) - \frac{h_v}{q_v}\delta q^{\rm m}_v(\bi{k})\right]\rangle^{[1]}=
 c_{\varepsilon}(z)\\
\label{6.17}
\fl 
\frac{\beta}{V}\,\langle\left[\frac{\delta q_v^{\rm m}(\bi{k})}{k}\right]^{\ast} 
\calL Q \frac{1}{z+Q\calL Q} Q\calL\, \left[\delta n_{\sigma}^{\rm
      m}(\bi{k}) - \frac{n_{\sigma}}{q_v}\delta q^{\rm m}_v(\bi{k})\right]\rangle^{[1]}=
 c_{\sigma}(z)\\
\label{6.18}
\fl
\frac{\beta}{\sqrt{m_v}\, V}\,\langle\left[\bi{g}^{\rm m}(\bi{k})\right]^{\ast} 
\calL Q \frac{1}{z+Q\calL Q} Q\calL \, 
\left[\delta\varepsilon^{\rm m}(\bi{k}) 
- \frac{h_v}{q_v}\delta q^{\rm m}_v(\bi{k})\right]\rangle^{[1]}\nonumber\\
=b_{\varepsilon}(z)\, \hat\bi{k}_{\perp}+b'_{\varepsilon}(z)\, 
\hat\bi{k}_{\perp}\wedge\hat\bi{B}\\
\label{6.19}
\fl
\frac{\beta}{\sqrt{m_v}\, V}\,\langle\left[\bi{g}^{\rm m}(\bi{k})\right]^{\ast} 
\calL Q \frac{1}{z+Q\calL Q} Q\calL \, 
\left[\delta n^{\rm m}_{\sigma}(\bi{k}) 
- \frac{n_\sigma}{q_v}\delta q^{\rm m}_v(\bi{k})\right]\rangle^{[1]}\nonumber\\
=b_{\sigma}(z)\, \hat\bi{k}_{\perp}+b'_{\sigma}(z)\, 
\hat\bi{k}_{\perp}\wedge\hat\bi{B}
\end{eqnarray}

The oscillating mode amplitudes $a^{{\rm m}(1)}_{\lambda\rho}$ up to
first order in the wave number have been obtained as the sum of (\ref{6.14})
and (\ref{6.15}). In the following we shall also need the thermal and
diffusive mode amplitudes up to first order in $k$. These have been derived
in \cite{GAQSLGS01} as 
\begin{eqnarray}
\label{6.20}
a^{{\rm m}(1)}_{\varepsilon}(\bi{k})=\delta\varepsilon^{\rm
  m}(\bi{k})-\frac{h_v}{q_v}\,\delta q^{\rm m}_v(\bi{k})
+k\, \frac{1}{\sqrt{m_v}}\, \bi{A}_{\varepsilon}^{\rm m}\cdot \bi{g}^{\rm 
m}(\bi{k}) \\
\label{6.21}
a^{{\rm m}(1)}_{\sigma}(\bi{k})=\delta n_{\sigma}^{\rm
  m}(\bi{k})-\frac{n_{\sigma}}{q_v}\,\delta q^{\rm m}_v(\bi{k})
+k\, \frac{1}{\sqrt{m_v}}\, \bi{A}_{\sigma}^{\rm m}\cdot \bi{g}^{\rm 
m}(\bi{k})
\end{eqnarray}
The coefficients $\bi{A}^{\rm m}_j$ have the decomposition
\begin{equation}
\label{6.22}
\bi{A}^{\rm m}_j=A^{{\rm m}\parallel}_j\, \hat\bi{k}_{\parallel}
+A^{{\rm m}\perp}_j\, \hat\bi{k}_{\perp}
+A^{{\rm m}t}_j \, \hat\bi{k}_{\perp}\wedge\hat\bi{B} 
\end{equation}
with
\begin{eqnarray}
\label{6.23}
\fl A_j^{{\rm m}\parallel}=\frac{1}{\omega_p \hat{k}^2_{\parallel}}
\left( \frac{\rmi}{\gamma}\,  b'_j\, \hat{k}^2_{\perp}-c_j\right)\quad , \quad 
A_j^{{\rm m}\perp}=\frac{\rmi}{\gamma}\frac{b_j b'-b'_j (\omega_p+b)}
{(\omega_p+b)^2+b'^2}
\nonumber \\
A_j^{{\rm m}t}=\frac{\rmi}{\gamma}\frac{b_j (\omega_p+b)+b'_j b'}
{(\omega_p+b)^2+b'^2}
\end{eqnarray}
for $j=\varepsilon , \sigma (\neq 1)$. The functions $b$, $b'$, $b_j$,
$b'_j$ and $c_j$ are given by (\ref{6.12}),
(\ref{6.16})--(\ref{6.19}). Here $z$ is equal to the infinitesimal value
$\rmi 0$, since the frequencies $z^{{\rm m}(0)}_j$ of the heat and
diffusion modes vanish in lowest order of $k$.

\section{Kinetic derivation of the frequencies in second order of the wave number}
\label{sec7}
\setcounter{equation}{0}
The frequencies $z^{\rm m[2]}_i$ in second order of the wave number can be
obtained without having to solve the full eigenvalue problem (\ref{6.9}) in
second order. In fact, it can be shown (see \ref{appB}) that knowledge of
the mode amplitudes up to first order is enough to evaluate the
second-order frequencies, since the latter follow as:
\begin{eqnarray}
\label{7.1}
\fl z^{{\rm m}[2]}_i 
=-z^{{\rm m}(0)}_i\,\frac{1}{V}
\langle \left[\bar a_i^{{\rm m}(1)}(\bi{k})\right]^{\ast}\, 
a^{{\rm m}(1)}_i (\bi{k})\rangle^{[2]}
-\frac{1}{V}\langle \left[\bar a_i^{{\rm m}(1)}(\bi{k})\right]^{\ast}\, 
\calL a^{{\rm m}(1)}_i (\bi{k})\rangle^{[2]}\nonumber\\
+\frac{1}{V}\langle \left[\bar a_i^{{\rm m}(1)}(\bi{k})\right]^{\ast}\, 
\calL Q\frac{1}{z^{{\rm m}(0)}_i+Q\calL
  Q} Q\calL a^{{\rm m}(1)}_i (\bi{k})\rangle^{[2]}
\end{eqnarray}
In this expression we have introduced the adjoint modes $\bar a^{\rm
m}_i(\bi k)$. These adjoints are specific independent linear combinations
of the basis functions $a^{\rm b}_i(\bi k)$ (\ref{6.6}). They are defined
by the orthonormality relations
\begin{equation}
\label{7.2}
\frac{1}{V}\langle [\bar a^{\rm m}_i(\bi k)]^{\ast} a^{\rm m}_j(\bi
k)\rangle = \delta_{ij}
\end{equation}

From equation (\ref{7.1}) it follows that the adjoints of the modes
(\ref{6.14})--(\ref{6.15}) up to first order in the wave number have to be
determined. In deriving these adjoints from (\ref{7.2}) it will be assumed
that the transport properties $b(z)$, $b'(z)$, $b_j(z)$, $b'_j(z)$ and
$c_j(z)$ are slowly varying functions of $z$, which can be treated as
constants. In zeroth order the adjoints $\bar a_{\lambda \rho}^{{\rm
m}(0)}(\bi{k})$ are given by the general expression \cite{AJSLGS90}
\begin{equation} 
\label{7.3}
\bar a_{\lambda \rho}^{{\rm m}(0)}(\bi{k}) = 
\beta(N^{\rm m}_{\lambda\rho})^{\ast}
\,\left[\frac{\delta q_v^{\rm m}(\bi{k})}{k}  +
\frac{1}{\sqrt{m_v}} \bar\bi{v}_{\lambda \rho}^{\rm m}(\bi{k}) \cdot 
\bi{g}^{\rm m}(\bi{k})\right]
\end{equation}
where the vector $\bar\bi{v}_{\lambda \rho}^{\rm m}(\bi{k})$ can be
written as
\begin{equation}
\label{7.4}
\bar\bi{v}_{\lambda\rho}^{\rm m}=
\bar v^{{\rm m}\parallel}_{\lambda\rho}\, \hat\bi{k}_{\parallel} 
+\bar v^{{\rm m}\perp}_{\lambda\rho}\, \hat\bi{k}_{\perp}\, 
+\bar v^{{\rm m}t}_{\lambda\rho}\, \hat\bi{k}_{\perp}\wedge\hat\bi{B}
\end{equation}
The normalization constant $N^{\rm m}_{\lambda\rho}$ and the coefficients $\bar
v^{{\rm m}\parallel}_{\lambda\rho}$, $\bar v^{{\rm m}\perp}_{\lambda\rho}$ and $\bar
v^{{\rm m}t}_{\lambda\rho}$ follow from the orthonormality relations in
zeroth order:
\begin{equation}
\label{7.5}
\frac{1}{V}\langle [\bar a_{\lambda\rho}^{{\rm m}(0)}(\bi{k})]^{\ast} 
a_{\mu\sigma}^{{\rm m}(0)}
(\bi{k}) \rangle^{(0)} = \delta_{\lambda \mu}\delta_{\rho \sigma}
\end {equation}
with $\lambda=\pm 1$, $\rho=\pm 1$, $\mu=\pm 1$ and $\sigma=\pm 1$. 
With the help of the fluctuation formulas \cite{AJSLGS90} these
orthonormality relations can be brought into the form
\begin{equation}
\label{7.6}
N^{\rm m}_{\lambda\rho}\left\{
1 + [\bar\bi{v}_{\lambda \rho}^{\rm m}(\bi k)]^{\ast} \cdot
\bi{v}_{\mu \sigma}^{\rm m}(\bi k)\right\} = \delta_{\lambda\mu}\delta_{\rho\sigma}
\end{equation}
The coefficients $\bar v^{{\rm m}\parallel}_{\lambda\rho}$, 
$\bar v^{{\rm m}\perp}_{\lambda\rho}$ and $\bar v^{{\rm m}t}_{\lambda\rho}$
may be obtained from this equality by choosing combinations of indices
$(\lambda\rho)$ that differ from $(\mu\sigma)$. In that case the
orthonormality relations read
\begin{equation}
\label{7.7}
\fl (\bar v^{{\rm m}\parallel}_{\lambda\rho})^{\ast}v^{{\rm
    m}\parallel}_{\mu\sigma}\hat{k}_{\parallel}^2
+(\bar v^{{\rm m}\perp}_{\lambda\rho})^{\ast}v^{{\rm
    m}\perp}_{\mu\sigma}\hat{k}_{\perp}^2
+(\bar v^{{\rm m}t}_{\lambda\rho})^{\ast}v^{{\rm
    m}t}_{\mu\sigma}\hat{k}_{\perp}^2 = -1
\quad , \quad (\lambda\rho)\neq (\mu\sigma)
\end{equation}
Let us now consider two different solutions $z^{{\rm m}(0)}_{\lambda \rho}$ and
$z^{{\rm m}(0)}_{\mu \sigma}$ of the quartic equation (\ref{4.3}). An
alternative form of the latter is
\begin{equation}
\label{7.8}
c - z^{{\rm m}(0)}_{\lambda\rho} = -\omega_p v^{{\rm
    m}\parallel}_{\lambda\rho}\hat{k}_{\parallel}^2
-(\omega_p+b)v^{{\rm m}\perp}_{\lambda\rho}\hat{k}_{\perp}^2
+b' v^{{\rm m}t}_{\lambda\rho}\hat{k}_{\perp}^2
\end{equation}
Subtraction of this equality from its analogue with
$(\lambda\rho)\rightarrow (\mu\sigma)$ and division by $z^{{\rm
m}(0)}_{\lambda\rho} - z^{{\rm m}(0)}_{\mu\sigma}$ yields the identity
\begin{equation}
\label{7.9}
v^{{\rm m}\parallel}_{\lambda\rho}\,v^{{\rm
    m}\parallel}_{\mu\sigma}\hat{k}_{\parallel}^2
+v^{{\rm m}\perp}_{\lambda\rho}\,v^{{\rm
    m}\perp}_{\mu\sigma}\hat{k}_{\perp}^2
-v^{{\rm m}t}_{\lambda\rho}\,v^{{\rm m}t}_{\mu\sigma}\hat{k}_{\perp}^2 = -1
\quad , \quad (\lambda\rho)\neq (\mu\sigma)
\end{equation}
By comparison with (\ref{7.7}) one infers that the coefficients $\bar
v^{{\rm m}\parallel}_{\lambda\rho}$, $\bar v^{{\rm m}\perp}_{\lambda\rho}$
and $\bar v^{{\rm m}t}_{\lambda\rho}$ are given by
\begin{equation}
\label{7.10}
\bar v^{{\rm m}\parallel}_{\lambda\rho} = 
(v^{{\rm m}\parallel}_{\lambda\rho})^{\ast}\quad , \quad 
\bar v^{{\rm m}\perp}_{\lambda\rho} =
(v^{{\rm m}\perp}_{\lambda\rho})^{\ast}\quad , \quad
\bar v^{{\rm m}t}_{\lambda\rho} = -(v^{{\rm m}t}_{\lambda\rho})^{\ast}
\end{equation}
The normalization constant $N^{\rm m}_{\lambda \rho}$ follows from
(\ref{7.6}) for $(\lambda\rho)=(\mu\sigma)$. It has the same structure as
(\ref{5.13}).

The contribution of first order in the adjoint $\bar a^{{\rm
m}[1]}_{\lambda\rho}(\bi{k})$ takes the general form
\begin{equation}
\label{7.11}
\bar a^{{\rm m}[1]}_{\lambda\rho}(\bi{k})= 
\beta\,(N^{\rm m}_{\lambda\rho})^{\ast}\,\left[
\bar A^{\rm m}_{\lambda\rho,\varepsilon}\, \delta\varepsilon^{\rm m}(\bi{k})
+ \sum_{\sigma(\neq 1)} \bar A^{\rm m}_{\lambda\rho,\sigma}\, 
\delta n_{\sigma}^{\rm m}(\bi{k})\right]
\end{equation}
The coefficients $\bar A^{\rm m}_{\lambda\rho,\varepsilon}$ and $\bar
A^{\rm m}_{\lambda\rho,\sigma}$ can be obtained from the orthogonality
relations
\begin{equation}
\label{7.12}
\frac{1}{V}\langle [\bar a_{\lambda \rho}^{{\rm m}(1)}(\bi{k})]^{\ast} a_{j}^{{\rm m}(1)}
(\bi{k}) \rangle^{(1)} = 0
\end {equation}
with $\lambda=\pm 1$, $\rho=\pm 1$ and $j=\varepsilon, \sigma (\neq
1)$. Substituting (\ref{7.3}), (\ref{7.11}) and (\ref{6.20}), (\ref{6.21})
into (\ref{7.12}) and employing the fluctuation formulas \cite{AJSLGS90},
\cite{AJVWLGS87} we get a set of inhomogeneous linear equations for the
unknown coefficients $\bar A^{\rm m}_{\lambda\rho,j}$. Using the
expressions (\ref{6.23}) for the coefficients $A^{{\rm m}\parallel}_j$,
$A^{{\rm m}\perp}_j$ and $A^{{\rm m}t}_j$ we obtain the solutions:
\begin{equation}
\label{7.13}
 \bar A^{\rm m}_{\lambda\rho,j}= 
\left(A^{\rm m}_{\lambda\rho,j}\right)^{\ast}
\end{equation}
with $j = \varepsilon, \sigma(\neq 1)$. 

Having succeeded in deriving the adjoints up to first order in $k$ we are
ready to evaluate the terms of second order $z^{{\rm m}[2]}_{\lambda\rho}$
in the mode frequencies. In the first term of (\ref{7.1}) we employ the
fluctuation formulas and the thermodynamic relations of \ref{appA} up to
second order in $k$. In particular, we use the Stillinger-Lovett relation
for the ionic mixture, as given in formula (6.6) of \cite{AJVWLGS87}, and
the closely connected fluctuation formula involving the pressure and the
charge density (see (7.9) of \cite{AJVWLGS87}). As a result we get from the
first term of (\ref{7.1}):
\begin{eqnarray}
\label{7.14}
\fl  
N^{\rm m}_{\lambda\rho}\; z^{{\rm m}(0)}_{\lambda\rho}\, \left(
-\beta \sum_{j,j'}A^{\rm m}_{\lambda\rho,j}\, M_{jj'}\, A^{\rm
  m}_{\lambda\rho,j'}
-2\sum_{j,j'}A^{\rm m}_{\lambda\rho,j}\,M_{jj'}\, s_{j'}
-2A^{\rm m}_{\lambda\rho,\varepsilon}\,\frac{4\varepsilon}{3q_v}\right.\nonumber\\
\left. -2\sum_{\sigma(\neq 1)}A^{\rm m}_{\lambda\rho,\sigma}\,
\frac{n_{\sigma}}{q_v}
-\frac{1}{\beta}\sum_{j,j'}s_j\, M_{jj'}\, s_{j'}
+\frac{4\varepsilon}{9q_v^2}+\frac{n}{2\beta q_v^2}\right)
\end{eqnarray}
A more useful form, which partly resembles (\ref{5.12}), is found by
employing (\ref{A11}) and the kinetic analogue of (\ref{4.22}):
\begin{eqnarray}
\label{7.15}
\fl  
-N^{\rm m}_{\lambda\rho}\; z^{{\rm m}(0)}_{\lambda\rho}
\left[ A^{\rm m}_{\lambda\rho,\varepsilon}\,\left(
F^{\rm m}_{\lambda\rho,\varepsilon}+\frac{h_v}{q_v}\right)
+\sum_{\sigma(\neq 1)}A^{\rm m}_{\lambda\rho,\sigma}\,\left(
F^{\rm m}_{\lambda\rho,\sigma}+\frac{n_{\sigma}}{q_v}\right)\right.
\nonumber\\
\left.+\frac{1}{\beta}\left(
\sum_j F^{\rm m}_{\lambda\rho,j}\, W_j+\frac{n}{2\beta q_v}\,
W_{\varepsilon}-\frac{1}{2q_v e_1}\right)\right]
\end{eqnarray}
In the second line we recognize the kinetic counterpart
$D^{\rm m}_{\lambda\rho,q}\!\mbox{}'$ of (\ref{5.14}).

In the second term of (\ref{7.1}) we insert the laws (\ref{6.3}) and
(\ref{6.4}). Using the fluctuation formulas once again, we find:
\begin{equation}
\label{7.16}
2 \, N^{\rm m}_{\lambda\rho}\;\omega_p\,
 \left( \frac{h_v}{q_v}\, A^{\rm m}_{\lambda\rho,\varepsilon}+
\sum_{\sigma(\neq 1)} \frac{n_{\sigma}}{q_v}\, A^{\rm m}_{\lambda\rho,\sigma}\right)\,
\hat{\bi{k}}\cdot \bi{v}_{\lambda\rho}^{\rm m}
\end{equation}

Finally, we have to evaluate the last term of (\ref{7.1}). Inserting (\ref{6.14}),
(\ref{6.15}), (\ref{7.3}) and (\ref{7.11}), we obtain a sum of several
contributions. The first of these is obtained upon substituting the
zeroth-order mode amplitudes (\ref{6.14}) and (\ref{7.3}) in
(\ref{7.1}). Using (\ref{6.3})--(\ref{6.4}) and the identities
$Q\delta q^{\rm m}_v(\bi{k})=0$ and $Q\bi{g}^{\rm m}(\bi{k})=0$, we get
\begin{eqnarray}
\label{7.17}
\fl \frac{\rmi}{T}\,  N^{\rm m}_{\lambda\rho}\, 
\left(\hat{\bi{k}}-\rmi\, \gamma\,
\bi{v}_{\lambda\rho}^{\rm m}\wedge\hat{\bi{B}}\right) \cdot
\left(\hat{\bi{k}}\cdot\bfsfL^{\rm m(c)}\cdot 
\hat{\bi{k}}\right)\cdot\left(\hat{\bi{k}}+\rmi\, \gamma\,
\bar{\bi{v}}_{\lambda\rho}^{{\rm m}\ast}\wedge\hat{\bi{B}}\right)
\nonumber\\
+\frac{1}{T\sqrt{m_v}}\,  N^{\rm m}_{\lambda\rho}\, 
\left(\hat{\bi{k}}-\rmi\, \gamma\,
\bi{v}_{\lambda\rho}^{\rm m}\wedge\hat{\bi{B}}\right) \cdot
\left(\hat{\bi{k}}\cdot\bfsfL^{\rm m(c)}_v\cdot 
\hat{\bi{k}}\right)\cdot\bar{\bi{v}}_{\lambda\rho}^{{\rm m}\ast}\nonumber\\
+\frac{1}{T\sqrt{m_v}}\,  N^{\rm m}_{\lambda\rho}\,
\bi{v}_{\lambda\rho}^{\rm m}
\cdot \left(\hat{\bi{k}}\cdot\bfsfL^{\rm m(c)}_{v'}
\cdot \hat{\bi{k}}\right)
\cdot\left(\hat{\bi{k}}+\rmi\, \gamma\,
\bar{\bi{v}}_{\lambda\rho}^{{\rm m}\ast}\wedge\hat{\bi{B}}\right)
\nonumber\\
-\frac{\rmi}{m_v}\, N^{\rm m}_{\lambda\rho}\, \bi{v}_{\lambda\rho}^{\rm m}
\cdot \left(\hat{\bi{k}}\cdot\bfsfeta^{\rm m}
\cdot \hat{\bi{k}}\right)
\cdot \bar{\bi{v}}_{\lambda\rho}^{{\rm m}\ast}
\end{eqnarray}
Here we introduced the microscopic equivalents of the transport quantities
$\bfsfL^{\rm (c)}_{\sigma\sigma'}$, $\bfsfL^{\rm (c)}_{\sigma v}$,
$\bfsfL^{\rm (c)}_{v\sigma}$ $(\sigma, \sigma' \neq 1)$ and
$\bfsfeta$. They are defined as
\begin{eqnarray}
\label{7.18}
\fl
\frac{1}{V}\,\langle\left[(\bi{g}^{{\rm m}}_{\sigma'}(\bi{k}))^j\right]^{\ast}
 Q \frac{1}{z+Q\calL Q} Q \, 
(\bi{g}^{{\rm m}}_{\sigma}(\bi{k}))^i\rangle^{[2]}
={\rm i} k_B\left(\hat\bi{k}\cdot\bfsfL^{\rm m(c)}_{\sigma\sigma'}(z)
\cdot\hat\bi{k}\right)^{ij}\\
\label{7.19}
\fl
\frac{1}{V}\,\langle
\left[(\hat\bi{k}\cdot\delta\bfsftau^{\rm m}(\bi{k}))^j\right]^{\ast} 
 Q \frac{1}{z+Q\calL Q} Q \, 
(\bi{g}^{{\rm m}}_{\sigma}(\bi{k}))^i\rangle^{[1]}
=k_B\left(\hat\bi{k}\cdot\bfsfL^{\rm m(c)}_{\sigma
  v}(z)\cdot\hat\bi{k}\right)^{ij}\\
\label{7.20}
\fl
\frac{1}{V}\,\langle\left[(\bi{g}^{{\rm m}}_{\sigma}(\bi{k}))^j\right]^{\ast} 
 Q \frac{1}{z+Q\calL Q} Q \, 
(\hat\bi{k}\cdot\bfsftau^{\rm m}(\bi{k}))^i\rangle^{[1]}
=k_B\left(\hat\bi{k}\cdot\bfsfL^{\rm m(c)}_{v \sigma}(z)
\cdot\hat\bi{k}\right)^{ij}\\
\label{7.21}
\fl
\frac{\beta}{V}\,
\langle\left[(\hat\bi{k}\cdot\delta\bfsftau^{\rm m}(\bi{k}))^j\right]^{\ast} 
 Q \frac{1}{z+Q\calL Q} Q \, 
(\hat\bi{k}\cdot\bfsftau^{\rm m}(\bi{k}))^i\rangle^{(0)}
=-{\rm i}\left(\hat\bi{k}\cdot\bfsfeta^{\rm m}(z)\cdot\hat\bi{k}\right)^{ij}
\end{eqnarray}
with $i, j = 1, 2, 3$. In (\ref{7.17}) we employed reduced transport
quantities that are defined in a way analogous to (\ref{3.6}), (\ref{3.7})
and (\ref{3.9}). All these quantities are to be taken at $z=z^{{\rm
m}(0)}_{\lambda\rho}$. The contractions of the vectors $\hat{\bi{k}}$ in
(\ref{7.17})--(\ref{7.21}) are to be taken with respect to the two inner 
indices of the fourth-rank tensors; the right-hand side of (\ref{7.18}),
for instance, equals $\rmi k_B \hat{k}^p\, (\bfsfL^{\rm
m(c)}_{\sigma\sigma'}(z))^{ipqj}\, \hat{k}^q$.  The expression (\ref{7.17})
can be rewritten by inserting (\ref{7.4}) with (\ref{7.10}). In this way we
get:
\begin{equation}
\label{7.22}
\rmi\, N^{\rm m}_{\lambda\rho} \left(
D^{\rm m}_{\lambda\rho,q}\!\mbox{}''+
v^{{\rm m}\parallel}_{\lambda\rho}\,
\hat{k}^2_{\parallel}\, D_{\lambda\rho}^{{\rm m}\parallel}
+v^{{\rm m}\perp}_{\lambda\rho}\,
\hat{k}_{\perp}^2\, D_{\lambda\rho}^{{\rm m}\perp}
- v^{{\rm m}t}_{\lambda\rho}\,
\hat{k}_{\perp}^2\, D_{\lambda\rho}^{{\rm m}t}\right)
\end{equation}
where the coefficients $D^{\rm m}_{\lambda\rho,q}\!\mbox{}''$,
$D_{\lambda\rho}^{{\rm m}\parallel}$, $D_{\lambda\rho}^{{\rm m}\perp}$ and
$D_{\lambda\rho}^{{\rm m}t}$ have the same form as their
magnetohydrodynamical counterparts (\ref{5.15})--(\ref{5.18}).

Further contributions from the last term of (\ref{7.1}) arise by inserting
either (\ref{6.14}) and (\ref{7.11}) or (\ref{6.15}) and
(\ref{7.3}). Employing (\ref{6.11})--(\ref{6.13}) and
(\ref{6.16})--(\ref{6.19}) we get:
\begin{eqnarray}
\label{7.23}
\fl 2 \, N^{\rm m}_{\lambda\rho}\; A^{\rm m}_{\lambda\rho,\varepsilon}
\left\{ c_{\varepsilon}+\frac{h_v}{q_v}\, c+
\bi{v}_{\lambda\rho}^{\rm m}\cdot\left[\left(b_{\varepsilon}+\frac{h_v}{q_v}\,
  b\right)
\hat{\bi{k}}_{\perp}-\left(b'_{\varepsilon}+\frac{h_v}{q_v}\, b'\right)
\hat{\bi{k}}_{\perp}\wedge\hat{\bi{B}}\right]\right\}\nonumber\\
\rule{-2.5cm}{0cm}+2 \, N^{\rm m}_{\lambda\rho}\, \sum_{\sigma(\neq 1)}
A^{\rm m}_{\lambda\rho,\sigma}
\left\{ c_{\sigma}+\frac{n_\sigma}{q_v}\, c+
\bi{v}_{\lambda\rho}^{\rm m}\cdot\left[\left(b_{\sigma}+\frac{n_\sigma}{q_v}\,
  b\right)\hat{\bi{k}}_{\perp}
-\left(b'_{\sigma}+\frac{n_\sigma}{q_v}\, b'\right)
\hat{\bi{k}}_{\perp}\wedge\hat{\bi{B}}\right]\right\}\nonumber\\
\mbox{}
\end{eqnarray}
where (\ref{7.10}) and (\ref{7.13}) have been used. As before, all
quantities $b$, $b'$, $c$, $b_j$, $b'_j$ and $c_j$ are taken for $z=z^{{\rm
m}(0)}_{\lambda\rho}$. By means of the kinetic analogue of (\ref{4.23}) we
may write $c_j$ in terms of $F^{\rm m}_{\lambda\rho,j}$. Furthermore, the
quartic equation (\ref{4.3}) can be used to eliminate $c$. As a result,
(\ref{7.23}) gets the form:
\begin{eqnarray}
\label{7.24}
2\, N^{\rm m}_{\lambda\rho}\; A^{\rm m}_{\lambda\rho,\varepsilon}
\left[z^{{\rm m}(0)}_{\lambda\rho}\left( F^{\rm m}_{\lambda\rho,\varepsilon}+
\frac{h_v}{q_v}\right)-\omega_p\,\frac{h_v}{q_v}\, 
\hat{\bi{k}}\cdot\bi{v}_{\lambda\rho}^{\rm m}\right]\nonumber\\
+2\, N^{\rm m}_{\lambda\rho}\, \sum_{\sigma(\neq 1)}
A^{\rm m}_{\lambda\rho,\varepsilon}
\left[ z^{{\rm m}(0)}_{\lambda\rho}\left(F^{\rm m}_{\lambda\rho,\sigma}+
\frac{n_\sigma}{q_v}\right)-\omega_p\,\frac{n_\sigma}{q_v}\, 
\hat{\bi{k}}\cdot\bi{v}_{\lambda\rho}^{\rm m}\right]
\end{eqnarray}

By adding (\ref{7.15}), (\ref{7.16}), (\ref{7.22}) and (\ref{7.24}) we find
an expression for the second-order frequencies that is very similar to
(\ref{5.12})
\begin{eqnarray}
\label{7.25}
\fl z^{{\rm m}[2]}_{\lambda\rho}=N^{\rm m}_{\lambda\rho}\,\left\{
z^{{\rm m}(0)}_{\lambda\rho}\,\left[ 
A^{\rm m}_{\lambda\rho,\varepsilon}\,
\left( F^{\rm m}_{\lambda\rho,\varepsilon}+\frac{h_v}{q_v}\right)+
\sum_{\sigma(\neq 1)}A^{\rm m}_{\lambda\rho,\sigma}\,
\left(F^{\rm m}_{\lambda\rho,\sigma}+\frac{n_{\sigma}}{q_v}\right)\right]\right.
\nonumber\\
\left.+\rmi\, D^{\rm m}_{\lambda\rho,q}+\rmi\,
v^{{\rm m}\parallel}_{\lambda\rho}\,
\hat{k}^2_{\parallel}\, D_{\lambda\rho}^{{\rm m}\parallel}
+\rmi\, v^{{\rm m}\perp}_{\lambda\rho}\,
\hat{k}_{\perp}^2\, D_{\lambda\rho}^{{\rm m}\perp}
-\rmi\, 
v^{{\rm m}t}_{\lambda\rho}\,
\hat{k}_{\perp}^2\, D_{\lambda\rho}^{{\rm m}t}\rule{0cm}{0.6cm}\right\}
\end{eqnarray}
Note that in this kinetic derivation the coefficients $A^{\rm
m}_{\lambda\rho,j}$ and $F^{\rm m}_{\lambda\rho,j}$ take the role of their
magnetohydrodynamical counterparts $A_{\lambda\rho,j}$ and $F_{\lambda\rho,j}$
(with $j = \varepsilon, \sigma (\neq 1)$). Likewise, the coefficients
$D_{\lambda\rho,q}^{\rm m}$, $D_{\lambda\rho}^{{\rm m}\parallel}$,
$D_{\lambda\rho}^{{\rm m}\perp}$ and $D_{\lambda\rho}^{{\rm m}t}$ replace
their magnetohydrodynamical equivalents (\ref{5.14})--(\ref{5.18}). Upon
substituting the kinetic analogues of (\ref{4.22}) in (\ref{7.25}) we
finally arrive at the kinetic expression for the second-order contributions
to the eigenfrequencies of the oscillating modes:
\begin{eqnarray}
\label{7.26}
\fl z^{{\rm m}[2]}_{\lambda\rho}=N^{\rm m}_{\lambda\rho}\,\left\{
z^{{\rm m}(0)}_{\lambda\rho}\,\left[\frac{c_s^2}{\omega^2_p} 
+\frac{2}{\beta}\sum_j\left(\frac{n}{2\beta q_v}\, M^{-1}_{\varepsilon
  j}-s_j\right) F^{\rm m}_{\lambda\rho,j}
+\frac{1}{\beta}\sum_{j,j'} F^{\rm m}_{\lambda\rho,j}\, M^{-1}_{jj'}\,
F^{\rm m}_{\lambda\rho,j'}
\right]\right.\nonumber\\
\left.
+\rmi\, D^{\rm m}_{\lambda\rho,q}\!\mbox{}''+\rmi\,v^{{\rm m}\parallel}_{\lambda\rho}\,
\hat{k}^2_{\parallel}\, D_{\lambda\rho}^{{\rm m}\parallel}
+\rmi\,v^{{\rm m}\perp}_{\lambda\rho}\,
\hat{k}_{\perp}^2\, D_{\lambda\rho}^{{\rm m}\perp}
-\rmi\, v^{{\rm m}t}_{\lambda\rho}\,
\hat{k}_{\perp}^2\, D_{\lambda\rho}^{{\rm m}t}\rule{0cm}{0.6cm}\right\}
\end{eqnarray}
This expression has the same form as (\ref{5.19}). Hence, the kinetic
approach corroborates the magnetohydrodynamical results for the damping and
the dispersion of the oscillating modes, at least up to second order in the wavelength.

\section{Conclusion}
\label{sec8}
\setcounter{equation}{0} In this paper we have shown how the spectrum of
collective modes of an ionic mixture with an arbitrary number of species
can be analyzed by means of magnetohydrodynamical and kinetic methods. We
have derived explicit expressions for the eigenfrequencies of the
oscillating modes up to second order. Since the second-order contributions
to the frequencies determine the damping and the dispersion of the
collective modes, our results give useful information on the dynamical
behaviour of ionic mixtures. As we have seen, the damping and dispersion
phenomena are governed by dissipative mechanisms that go beyond those of
standard non-equilibrium thermodynamics. In fact, the well-known
phenomenological laws for the diffusive flows and the viscous pressure are
not sufficient to account for the full effects of dispersion and damping of
the oscillating modes. Whereas the dissipation implied by these standard
laws yields the damping and dispersion of the heat mode and the
diffusion modes, as shown in our previous paper \cite{GAQSLGS01}, we had to
generalize the standard laws by including higher-order Burnett terms in
order to get a complete picture of the damping and the dispersion of the
oscillating modes up to second-order in the wave number.

Burnett terms have been discussed previously in the context of systems of
neutral particles \cite{DB36}--\cite{JJB81}. For such systems the Burnett
terms generally lead to small effects in the dynamical behaviour, so that
they can usually be neglected. In the present case of an ionic mixture that
is no longer true: in a systematic treatment of the collective modes the
Burnett terms can not be disregarded. The technical reason for the
necessity to incorporate Burnett terms when investigating the dynamics of
an ionic mixture turns out to be the long-range nature of the Coulomb
interactions. In an ionic mixture the long-range electric fields generated
by charge fluctuations are driving agents in the thermodynamical
forces. These electric fields lead to terms of zeroth order in the
wavenumber after Fourier transformation. To be consistent up to second
order in the wavenumber one has to include terms with gradients of the
electric field in the thermodynamic forces. Hence the phenomenological
laws, which connect these thermodynamical forces to the thermodynamic
flows, must be used in a form which accounts for such field gradient
terms. We have to conclude that Burnett terms can not be neglected in
systems with long-range forces.

\ack

One of the authors (GAQS) would like to acknowledge the financial support
of the `Nederlandse Organisatie voor Wetenschappelijk Onderzoek (NWO)' as
part of its program `Leraar in Onderzoek'.

\appendix 
\section{Thermodynamics for an ionic mixture}
\label{appA}
An ionic mixture satisfies the global constraint of charge
neutrality. Therefore, it is convenient to introduce instead of
$\mu_{\sigma}$ a set of new chemical potentials:
\begin{equation}
\label{A1}
\tilde{\mu}_{\sigma}=\mu_{\sigma}-\frac{e_{\sigma}}{e_1}\, \mu_1 
\end{equation}
for $\sigma\neq 1$. As shown in \cite{LGSAJVW87}, these potentials arise
naturally as Lagrange multipliers in a grand-canonical ensemble with a
charge-neutrality constraint. The combination of perturbations of the
chemical potentials $\mu_{\sigma}$ occurring in the conservation laws and
the balance equations of section \ref{sec3} can be rewritten in terms of
these new chemical potentials as \cite{GAQSLGS01}:
\begin{eqnarray}
\label{A2}
\fl \left(\frac{\delta\mu_{\sigma}}{m_{\sigma}}-
\frac{\delta\mu_1}{m_1}\right)_T=
-T\left[\frac{h_{\sigma}}{m_{\sigma}}-\frac{h_1}{m_1}
-  \left(\frac{e_{\sigma}}{m_{\sigma}}-\frac{e_1}{m_1}\right)\, 
\frac{h_v}{q_v}\right]\, \delta\left(\frac{1}{T}\right)
\nonumber\\
+T\sum_{\sigma'(\neq 1)}
\left[\frac{1}{m_{\sigma'}}\delta_{\sigma\sigma'}
-\left(\frac{e_{\sigma}}{m_{\sigma}}-\frac{e_1}{m_1}\right)\, 
\frac{n_{\sigma'}}{q_v}\right]\, 
\delta\left(\frac{\tilde{\mu}_{\sigma'}}{T}\right)
\nonumber\\
+\left(\frac{e_{\sigma}}{m_{\sigma}}-\frac{e_1}{m_1}\right)
\,\frac{\delta p}{q_v}
\end{eqnarray}
for $\sigma\neq 1$.  

The perturbations $\delta(1/T)$ and $\delta(\tilde{\mu}_{\sigma'}/T)$ (with
$\sigma'\neq 1$) in (\ref{A2}) have to be represented as linear combinations
of the basic perturbations $\delta\varepsilon$, $\delta n_{\sigma}$ and
$\delta q_v$. This can be achieved by starting from the identity \cite{GAQSLGS01}
\begin{equation}
\label{A3}
\left( \begin{array}{c}
\delta \varepsilon \\ \delta n_{\sigma} \end{array}\right) 
= \bfsfM \cdot \left( \begin{array}{c}
-\delta \beta \\ \delta (\beta \tilde{\mu}_{\sigma}) \end{array} \right)
+ \bi{V} \, \delta q_v 
\end{equation}
with the matrix:
\begin{equation}
\label{A4}
\bfsfM=\left( \begin{array}{cc}
-\partial\varepsilon/\partial\beta &  -\partial n_{\sigma'}/\partial\beta\\
-\partial n_{\sigma}/\partial\beta & 
\partial n_{\sigma}/\partial(\beta\tilde{\mu}_{\sigma'})
\end{array} \right)
\end{equation}
and the vector
\begin{equation}
\label{A5}
\bi{V}=\left( \begin{array}{c}
\partial\varepsilon/\partial q_v \\
\partial n_{\sigma}/\partial q_v \end{array} \right)
\end{equation}
All partial derivatives are defined in terms of the independent variables
$\beta=1/(k_B T)$, $\beta\tilde{\mu}_{\sigma}$ (with $\sigma\neq 1$) and
$q_v$. We used the Maxwell relation
$\partial\varepsilon/\partial(\beta\tilde{\mu}_{\sigma})=-\partial
n_{\sigma}/\partial\beta$. 

The elements of $\bi{V}$ can be evaluated by substituting the equation of
state $p=\frac{1}{3}\varepsilon+\frac{1}{2}nk_B T$ into the formulas (3.22)
and (3.18) of ref. \cite{LGSAJVW87}, and replacing $n$ by
$-\sum_{\sigma(\neq 1)}(e_{\sigma}/e_1-1)\, n_{\sigma}+q_v/e_1$. Treating
$\beta$, $\beta\tilde\mu_{\sigma}$ ($\sigma \neq 1$) and $q_v$ as
independent variables, we get
\begin{equation}
\label{A6}
\frac{\partial\varepsilon}{\partial q_v} = 
M_{\varepsilon\varepsilon}\, s_{\varepsilon}+
\sum_{\sigma (\neq 1)}M_{\varepsilon \sigma}s_{\sigma} 
+ \frac{4\varepsilon}{3q_v}
\end{equation}
\begin{equation}
\label{A7}
\frac{\partial n_{\sigma}}{\partial q_v} = 
M_{\sigma\varepsilon}s_{\varepsilon}+
\sum_{\sigma' (\neq 1)}M_{\sigma\sigma'}s_{\sigma'} 
+ \frac{n_{\sigma}}{q_v}
\end{equation}
with $\sigma\neq 1$, and with the abbreviations:
\begin{equation}
\label{A8}
s_{\varepsilon} = -\frac{\beta}{3q_v} \quad , \quad  
s_{\sigma}=\frac{1}{2q_v}\left(\frac{e_{\sigma}}{e_1}-1\right)
\end{equation}
Inverting (\ref{A3}) and substituting (\ref{A6}) and (\ref{A7}) we finally
get:
\begin{eqnarray}
\label{A9}
-\delta\beta = M^{-1}_{\varepsilon\varepsilon}\delta\varepsilon + 
\sum_{\sigma (\neq 1)}M^{-1}_{\varepsilon\sigma}\delta n_{\sigma}
-k\, W_{\varepsilon}\, \frac{\delta q_v}{k}
\end{eqnarray}
\begin{eqnarray}
\label{A10}
\delta(\beta\tilde\mu_{\sigma}) = M^{-1}_{\sigma\varepsilon}\delta\varepsilon + 
\sum_{\sigma' (\neq 1)}M^{-1}_{\sigma\sigma'}\delta n_{\sigma'}
-k\, W_{\sigma}\, \frac{\delta q_v}{k}
\end{eqnarray}
with $W_j$ defined as
\begin{equation}
\label{A11}
W_j=\sum_{j'}M^{-1}_{jj'}V_{j'}=
s_{j}+M^{-1}_{j\varepsilon}\frac{4\varepsilon}{3q_v}+\sum_{\sigma'
  (\neq 1)}M^{-1}_{j\sigma'}\frac{n_{\sigma'}}{q_v}
\end{equation}
for $j$ and $j'$ equal to $\varepsilon$ or $\sigma (\neq 1)$. 

The variation in the pressure $\delta p$ can likewise be expressed in
terms of variations $\delta\varepsilon$, $\delta n_{\sigma}$ and $\delta
q_{v}/k$. In fact, the perturbation of the pressure follows from the
equation of state as:
\begin{equation}
\label{A12}
\fl \delta p=\frac{1}{3}\, \delta\varepsilon
-\frac{1}{2\beta}\sum_{\sigma(\neq 1)}
\left(\frac{e_{\sigma}}{e_1}-1\right)\, \delta n_{\sigma}
-\frac{n}{2\beta^2}\, \delta\beta
+\frac{1}{2e_1\beta}\, \delta q_v
\end{equation}
Upon inserting the expression (\ref{A9}) for $\delta\beta$ we obtain
\begin{eqnarray}
\label{A13}
\fl \beta\frac{\delta p}{q_v} 
= \left(\frac{n}{2\beta q_v}\, M^{-1}_{\varepsilon\varepsilon}
-s_{\varepsilon}\right)\delta\varepsilon
+\sum_{\sigma (\neq 1)}\left(\frac{n}{2\beta q_v}\, 
M^{-1}_{\varepsilon\sigma}-s_{\sigma}\right)
\delta n_{\sigma}\nonumber\\
+k\left(\frac{1}{2q_v e_1}-\frac{n}{2\beta q_v}\, W_{\varepsilon}\right)
\frac{\delta q_v}{k}
\end{eqnarray}

The inverse matrix element $M^{-1}_{\varepsilon\varepsilon}$ is directly related to
the specific heat $c_v$ per particle:
\begin{equation}
\label{A14}
c_v=T\left(\frac{\partial s}{\partial T}\right)_{q_v,\{ n_{\sigma}/q_v\}}
=\frac{\beta^2}{n\, M^{-1}_{\varepsilon\varepsilon}}\, k_B
\end{equation}
with $s$ the entropy per particle. Likewise, one may derive that the
specific heat $c_p$ per particle at constant pressure is given as
\begin{equation}
\label{A15}
c_p=T\left(\frac{\partial s}{\partial T}\right)_{p,\{n_{\sigma}/q_v\}}=
\frac{9\, n^2\, M^{-1}_{\varepsilon\varepsilon}+
2\,\beta^2(8\, \beta\,\varepsilon +15\, n)}
{2\, n(8\,\beta\,\varepsilon+9\, n)M^{-1}_{\varepsilon\varepsilon}
-4\, n\,\beta^2}\, k_B
\end{equation}
Furthermore, the isothermal compressibility $\kappa_T$ can be introduced by
any of the following equivalent definitions:
\begin{equation}
\label{A16}
\fl \kappa_T=\frac{1}{q_v}
\left(\frac{\partial q_v}{\partial p}\right)_{T,\{n_{\sigma}/q_v\}}
=\frac{1}{n}
\left(\frac{\partial n}{\partial p}\right)_{T,\{n_{\sigma}/q_v\}}=
\frac{1}{m_v}
\left(\frac{\partial m_v}{\partial p}\right)_{T,\{n_{\sigma}/q_v\}}
\end{equation}
where use has been made of the identity $dn_{\sigma}=(n_{\sigma}/q_v)\,
dq_v$ at fixed concentrations $n_{\sigma}/n$ (or fixed
$n_{\sigma}/q_v$). In terms of the inverse matrix element
$M^{-1}_{\varepsilon\varepsilon}$ the compressibility reads:
\begin{equation}
\label{A17}
\kappa_T=\frac{18\, \beta\,
    M^{-1}_{\varepsilon\varepsilon}}
{(8\,\beta\,\varepsilon +9\, n)M^{-1}_{\varepsilon\varepsilon}-2\, \beta^2}
\end{equation}
Combining these results one finds the sound velocity as
\begin{equation}
\label{A18}
c_s^2=\frac{c_p}{m_v\, c_v\, \kappa_T}=\frac{n^2}{4\, \beta^3\,m_v}\,
M^{-1}_{\varepsilon\varepsilon}+\frac{4\, \varepsilon}{9\, m_v}+
\frac{5\, n}{6\,\beta\, m_v}
\end{equation}
Alternatively, the right-hand side may be written in terms of the enthalpy
per volume $h_v=\frac{4}{3}\varepsilon+\frac{1}{2}n/\beta$.

\section{The eigenvalue problem in kinetic theory}
\label{appB}
\setcounter{equation}{0}

In the main text it has been stated that the frequencies $z^{\rm m}_i (\bi
k)$ and their corresponding modes $a^{\rm m}_i (\bi k)$ follow from the
eigenvalue problem (\ref{6.9}) in arbitrary order of $k$. However, for our
purposes it is more convenient to introduce adjoint modes $\bar
a^{\rm m}_i (\bi k)$. These adjoints are particular independent linear 
combinations of the basis functions $a^{\rm b}_i (\bi k)$. The adjoints
follow from the orthonormality condition
\begin{equation}
\label{B1}
\frac{1}{V}\langle [\bar a^{\rm m}_i(\bi k)]^{\ast} a^{\rm m}_j
(\bi k)\rangle = \delta_{ij}
\end{equation}
With the use of these adjoints, the eigenvalue problem (\ref{6.9}) up to
second order in the wave number may be written as 
\begin{eqnarray}
\label{B2}
\fl z^{{\rm m}(2)}_i\delta_{ij}=
-\frac{1}{V}\langle \left[\bar a_i^{{\rm m}(2)}(\bi{k})\right]^{\ast}\, 
\calL a^{{\rm m}(2)}_j (\bi{k})\rangle^{(2)}\nonumber\\
+\frac{1}{V}\langle \left[\bar a_i^{{\rm m}(2)}(\bi{k})\right]^{\ast}\, 
\calL Q\frac{1}{z^{{\rm m}(2)}_i+Q\calL
  Q} Q\calL a^{{\rm m}(2)}_j (\bi{k})\rangle^{(2)}
\end{eqnarray}
In this expression we have made explicit, through the notation between
round brackets in the superscripts, that the thermal averages
$\langle\ldots\rangle$ have to be evaluated up to second order in the wave
number. Contrary to what one might expect from the above expression, the
modes and their adjoints up to {\em first} order in $k$ are sufficient to
obtain the frequencies up to {\em second} order. This will be demonstrated
below.

Let us expand the second-order modes $a^{{\rm m}(2)}_i (\bi{k})$ in terms
of their first-order counterparts $a^{{\rm m}(1)}_i (\bi{k})$. Formally
this expansion can be written as
\begin{equation}
\label{B3}
a^{{\rm m}(2)}_i (\bi{k})=\sum_j C_{ij}a^{{\rm m}(1)}_j (\bi{k})
\end{equation}
where the coefficients $C_{ij}$ have the general form
\begin{equation}
\label{B4}
C_{ij}= \delta_{ij}+k^2\,C^{[2]}_{ij}
\end{equation}
In the same way we can expand the adjoints:
\begin{equation}
\label{B5}
\bar a^{{\rm m}(2)}_i (\bi{k})=\sum_j D_{ij}\bar a^{{\rm m}(1)}_j (\bi{k})
\end{equation}
with
\begin{equation}
\label{B6}
D_{ij}= \delta_{ij}+k^2\,D^{[2]}_{ij}
\end{equation}
The modes $a^{{\rm m}(1)}_i (\bi{k})$ and their adjoints are orthogonal up
to first order in the wave number, so that one has:
\begin{equation}
\label{B7}
\frac{1}{V}\langle [\bar a^{{\rm m}(1)}_i(\bi k)]^{\ast} a^{{\rm
    m}(1)}_j(\bi k)\rangle^{(2)} = \delta_{ij}+k^2\,F^{[2]}_{ij} 
\end{equation}
From the orthogonality of the modes $a^{{\rm m}(2)}_i (\bi{k})$ and their
adjoints in second order it then follows that the coefficients
$F^{[2]}_{ij}$ obey the relation
\begin{equation}
\label{B8}
F^{[2]}_{ij}=-(D^{[2]}_{ij})^{\ast}-C^{[2]}_{ji}
\end{equation}

Substitution of the expansions (\ref{B3}) and (\ref{B5}) into the
eigenvalue equation (\ref{B2}) yields up to second order in $k$:
\begin{eqnarray}
\label{B9}
\fl z^{{\rm m}(2)}_i\delta_{ij}=
\sum_{k,l} D^{\ast}_{ik}C_{jl}\left\{-\frac{1}{V}\langle \left[\bar
  a_k^{{\rm m}(1)}(\bi{k})\right]^{\ast}\,\calL a^{{\rm m}(1)}_l
(\bi{k})\rangle^{(2)}\right.\nonumber\\
\left.+\frac{1}{V}\langle \left[\bar a_k^{{\rm m}(1)}(\bi{k})\right]^{\ast}\, 
\calL Q\frac{1}{z^{{\rm m}(2)}_i+Q\calL
  Q} Q\calL a^{{\rm m}(1)}_l (\bi{k})\rangle^{(2)}\right\}
\end{eqnarray}
Upon approximating the frequencies $z^{{\rm m}(2)}_i$ by $z^{{\rm m}(0)}_i$
in the denominator of the second term at the right-hand side, we may write the
expression between curly brackets as
\begin{eqnarray}
\label{B10}
z^{{\rm m}(1)}_i\delta_{kl}+k^2\,G^{[2]}_{kl}\equiv 
-\frac{1}{V}\langle \left[\bar
  a_k^{{\rm m}(1)}(\bi{k})\right]^{\ast}\,\calL a^{{\rm m}(1)}_l
(\bi{k})\rangle^{(2)}\nonumber\\
+\frac{1}{V}\langle \left[\bar a_k^{{\rm m}(1)}(\bi{k})\right]^{\ast}\, 
\calL Q\frac{1}{z^{{\rm m}(0)}_i+Q\calL
  Q} Q\calL a^{{\rm m}(1)}_l (\bi{k})\rangle^{(2)}
\end{eqnarray}
which defines $G^{[2]}_{kl}$. Inserting (\ref{B10}) in (\ref{B9}) and using
the relations (\ref{B4}), (\ref{B6}) and (\ref{B8}), we obtain for 
$i=j$:
\begin{equation}
\label{B11}
z^{{\rm m}[2]}_i=-z^{{\rm m}(0)}_i F^{[2]}_{ii}+G^{[2]}_{ii}
\end{equation}
or, with (\ref{B7}) and again (\ref{B10})
\begin{eqnarray}
\label{B12}
\fl z^{{\rm m}[2]}_i=
-z^{{\rm m}(0)}_i\,\frac{1}{V}
\langle \left[\bar a_i^{{\rm m}(1)}(\bi{k})\right]^{\ast}\, 
a^{{\rm m}(1)}_i (\bi{k})\rangle^{[2]}
-\frac{1}{V}\langle \left[\bar a_i^{{\rm m}(1)}(\bi{k})\right]^{\ast}\, 
\calL a^{{\rm m}(1)}_i (\bi{k})\rangle^{[2]}\nonumber\\
+\frac{1}{V}\langle \left[\bar a_i^{{\rm m}(1)}(\bi{k})\right]^{\ast}\, 
\calL Q\frac{1}{z^{{\rm m}(0)}_i+Q\calL
  Q} Q\calL a^{{\rm m}(1)}_i (\bi{k})\rangle^{[2]}
\end{eqnarray}
From this expression for the eigenvalue problem it is obvious that the
modes and their adjoints up to first order in the wave number are
sufficient to determine the second-order contributions to the frequencies.

\end{document}